\documentclass[letterpaper,12pt]{article}  

\usepackage[round]{natbib}
\usepackage{amsmath}
\usepackage{amssymb}
\usepackage{amsthm}
\usepackage[USenglish]{babel}
\usepackage[utf8]{inputenc}
\usepackage[T1]{fontenc}
\usepackage{color}
\usepackage{graphicx}
\usepackage{enumerate}
\usepackage{ulem}
\usepackage{dcolumn}
\usepackage{multirow}
\usepackage{url}
\usepackage{colortbl}
\usepackage{authblk}
\usepackage[ruled,vlined]{algorithm2e}

\newlength{\textlarg}

\definecolor{darkred}{rgb}{.8,.0,.0}
\definecolor{gris}{rgb}{0.75,0.75,0.75}
\definecolor{lightgreen}{rgb}{.7,.95,.7}

\newcommand{\modif}{\color{black}}

\begin{document}

\title{RankMerging: A supervised learning-to-rank framework to predict links in large social networks}


\author[1]{Lionel Tabourier\thanks{corresponding author: \texttt{lionel.tabourier@ens-lyon.org}}}
\author[2]{Daniel F. Bernardes}
\author[3]{Anne-Sophie Libert}
\author[4]{Renaud Lambiotte}

\affil[1]{LIP6, UMR 7606, Sorbonne Université / CNRS, Paris, France}
\affil[2]{L2TI, Université Paris-Nord, France}
\affil[3]{naXys, University of Namur, Namur, Belgium}
\affil[4]{Mathematical Institute, University of Oxford, Oxford, UK}

\date{}

\maketitle

\begin{abstract}

Uncovering unknown or missing links in social networks is a difficult task because of their sparsity and because links may represent different types of relationships, characterized by different structural patterns. 
In this paper, we define a simple yet efficient supervised learning-to-rank framework, called \textit{RankMerging}, which aims at combining information provided by various unsupervised rankings. 
We illustrate our method on three different kinds of social networks and show that it substantially improves the performances of unsupervised methods of ranking as well as standard supervised combination strategies.
We also describe various properties of \textit{RankMerging}, such as its computational complexity, its robustness to feature selection and parameter estimation and discuss its area of relevance: the prediction of an adjustable number of links on large networks.

\end{abstract}


\section{Introduction} 

Link prediction is a key field of research for the mining and analysis of large-scale
social networks because of its many practical applications: going from
recommendation strategies for commercial websites~\citep{huang2005link} to recovering missing
links in incomplete data~\citep{zhou2009predicting}.
Link prediction also has significant implications
from a fundamental point of view, as it allows for the identification of the elementary mechanisms behind the creation and decay of links in time-evolving networks~\citep{leskovec2008microscopic}. 
For example, triadic closure, at the core of standard methods of
link prediction is considered as one of the driving forces for the creation of links
in social networks~\citep{kossinets2006empirical}.

In general, link prediction consists in inferring the existence of a set of links from the observed structure of a network.
The edges predicted may correspond to links that are bound to appear in the future, as in the seminal formulation by~\cite{liben2007link}.
They may also be existing links that have not been detected during the data collection process, in which case it is sometimes referred to as the missing link problem.  
In both cases, it can be described as a binary classification issue, where it is decided if a pair of nodes is connected or not.
The features used are often based on the structural properties of the network of known interactions, either at a local scale (e.g. the number of common neighbors) or at a global scale (e.g. random walk or hitting time).
See for example \cite{lu2011link} or \cite{al2011survey} for surveys. 
Other sources of information are available to predict links, in particular node attributes such as
age, gender or other profile information~\citep{backstrom2011supervised,bliss2013evolutionary}, geographic location~\citep{scellato2011exploiting}, as well as interaction attributes: frequencies~\citep{tylenda2009towards} or the time elapsed since the last interaction~\citep{raeder2011predictors}.

We consider this problem in the context of large social networks.
In this case, the classification issue has specific characteristics: for a typical $10^6$ nodes network, there are around $10^{12}$ candidate pairs of nodes that can be connected, most of them being completely irrelevant.
The problem is unmanageable without restraining ourselves to subsets of pairs.
{\modif
One way of doing so is to limit the prediction to pairs of nodes which are close enough, as there is indeed a higher probability that an edge appears between nodes located at short distance.}
Even with this restriction we have to handle typically rankings with $10^5$ to $10^7$ items, as we shall see it implies solving challenges in terms of computational efficiency.
Notice also that among these items, only a very small fraction are actually connected pairs, meaning that the two classes have very different sizes.
This problem is known as class imbalance and has an important impact on the link prediction -- see for example \cite{lichtenwalter2010new}.
We also discuss this question later in this paper.

%
%
The features used for classification are known to be domain-specific.
As links play various roles in social networks, they are expected to be surrounded by different types of environments and thus, to be best identified by different features.
For these reasons, schemes based on a single metric are prone to misclassification.
Machine learning methods have been widely used to combine the available information for the purpose of classification.
In recent works, classification trees, support vector machines, matrix factorization or neural networks are implemented to predict links in biological networks or scientific collaboration networks~\citep{pavlov2007finding,kashima2009link,benchettara2010supervised,lichtenwalter2010new,menon2011link,davis2013supervised}. 
However, these classification methods are not designed to easily set the number of predictions, while this property is highly desirable in the context of link prediction in social networks.
{\modif
Indeed, in many practical cases, we would like to set the number of predictions to a given value; for example if link prediction is used for recommendation purposes, a user might want to make a few high-precision predictions or on the contrary make a large number of predictions in order to reach all potential targets, depending on his or her strategy.
%
%
On the other hand, it means that the user has to set the number of predictions, which is not necessary with a classification method.}
In this work, we follow an approach which allows to easily set the number of predictions.

Another way to address the issue consists in establishing a ranking of likely links according to a scalar metric, correlated with the existence of interactions between nodes.
Therefore, we can use the same ranking features as those raised previously for the classification task, that is to say based on the structure or on node and link attributes.
Then, the user may set the number of links predicted by selecting the \textit{$ \theta $} top-ranked items.
Using this approach, the information brought by the various ranking features can be combined using \textit{learning-to-rank} frameworks.
The general idea is to aggregate these rankings in such a way that it creates a new, \textit{better}, output ranking.
Unsupervised solutions are available, such as Borda’s method or Markov chain ordering~\citep{dwork2001rank,sculley2007rank}.
These methods stem from social choice theory, where there is in general no ground truth available.
As a consequence, the merged ranking is built in order to yield the best consensus among the input rankings, defining a notion of distance of the output to the inputs.
In the situation of link prediction, it is possible to define a ground truth and therefore, to formulate the learning-to-rank task in a supervised way.

Supervised learning-to-rank techniques have been mostly designed in the context of information retrieval tasks, such as document filtering, spam webpage detection, recommendation or text summarization, see for example~\cite{freund2003efficient,liu2007supervised,burges2011learning,comar2011linkboost}.
In this field, the ground truth is the relevance evaluation of experts.
In~\cite{liu2009learning}, the author distinguishes between three kinds of approaches.
Firstly, pointwise approaches which are the most straightforward, using the score or rank associated to a feature to fit, for example, a regression model.
One undesirable effect is that low-ranked items tend to have an over-important role in the learning process, which is particularly critical in the case of link prediction as rankings are very large. 
Secondly, pairwise approaches~\citep{herbrich1999large} consist in transforming the ranking problem into a classification one, by considering couple of items and learning which one should be ranked above.
This transform allows the use of supervised classification methods, as the issue now consists in predicting if a couple of items $ (a,b) $ should be put in class $a$ over $b$ or in class $b$ over $a$.
Unfortunately even the cheapest implementations of this approach~\citep{chapelle2010efficient} cannot be used to predict links on large networks, as the number of items to rank here is larger than $ 10^5$.
Thirdly, listwise approaches~\citep{cao2007learning} use a ranking of items as ground truth.
This method is not relevant to our case because when predicting \textit{$\theta$} links, the quality of two rankings is strictly equivalent if they provide the same amount of true prediction in their top-\textit{$\theta$} items.
More generally, information retrieval techniques primarily aim at high precision on the top-ranked items, and stress the relative ranking of two items.
As stated in~\cite{chapelle2011future}, most of the research on the topic has therefore focused on improving the prediction accuracy rather than making the algorithms scalable, which is crucial in the case of link prediction.

Learning-to-rank in the context of link prediction in large graphs calls for specific methods, suited for large rankings.
In this spirit, \cite{subbian2011supervised} designed supervised ranking aggregation methods based on the adaptation of unsupervised methods.
However, they were created for another kind of prediction task, namely influence prediction.
Closer to our work, \cite{pujari2012supervised} adapted these methods to predict links in social networks using supervised approaches.
One method that they consider is based on giving weights to the Borda method according to the performances of the unsupervised rankings, in the following we refer to this aggregation method as \textit{Weighted Borda}.
Another, \textit{Kemeny aggregation}, is also inspired by methods trying to provide a consensus from the unsupervised rankings, by minimizing the number of pairwises disagreements between rankers.
This allowed efficient predictions on $10^4$ to $10^5$ items rankings.

In this work, we propose a simple yet efficient learning-to-rank supervised framework specifically designed to uncover links in large and sparse networks, such as social networks.
We improve the prediction by combining rankings obtained from different sources of information.
The article is organized as follows.
Section~\ref{sec:eval} is dedicated to the description of the features and metrics that we use to evaluate the performances of the link prediction.
We then present in Section~\ref{sec:unsupervised} how classic unsupervised learning methods can be applied to the problem under consideration.
In Section~\ref{sec:supervised}, we present our supervised machine learning framework, called \textit{RankMerging}, which improves the quality of predictions by aggregating the information from the unsupervised metrics.
Finally, we implement this method on three large social network datasets, during two series of experiments in Section~\ref{sec:exp}, and compare our results to those of other methods.
We also explore aspects such as the feature selection problem, the impact of parameters values etc, and show that \textit{RankMerging} is suited to social networks where information is partial and noisy, and the number of links to predict is large.


\section{Aggregation task and performance evaluation}

\label{sec:eval}

{\modif

\subsection{Description of the aggregation task}

In this work, link prediction is formulated as a learning-to-rank problem.
We describe here the rank aggregation task in general.
Suppose we have a set of items $ I = \{ item_1 , item_2 , \ldots , item_{max} \} $.
We also have $ \alpha $ lists, called rankings, which contain the whole set or a subset of $I$ in a definite order, for example $r_i = [r_i[1], r_i[2], \ldots , r_i[size_i]]$.
A ranking is \textit{full} if it contains all the elements of $I$ and \textit{partial} if it does not.
The aggregation task consists in building an output (or aggregated) ranking $r_o$ containing once all items of $I$.
In the context of link prediction, an item ranked is a pair of nodes and we build the output ranking in the purpose of discovering connected pairs at the top of $r_o$.}

\subsection{Performance evaluation for link prediction}

The definition of a \textit{good} aggregation and thus of an adequate quality estimator depends on the purpose of the ranking problem.
For example in the context of social choice theory, a ranking represents the order of preferences of a voter, so that the aggregation process aims at providing the best possible compromise of the different input rankings.
In this case, there is no ground truth and the quality of an aggregated ranking is often evaluated using a distance to the input rankings.
A usual choice is the Kendall tau distance between the input rankings $r_i$ and the output ranking $r_o$, which is defined as the number of couples of items which are ranked in different orders in $r_i$ and $r_o$. 
Notice that it is also true in the field of information retrieval, where the notion of goodness of a ranking is usually seen as a consensus among experts.

However considering link prediction, it is possible to define a ground truth on a learning set, where a link either exists or not.
It allows to use more appropriate performance metrics.
Considering a fixed number of predictions $\theta$, the link is predicted or not depending on whether its rank falls above or below $\theta$.
The quality of a prediction is therefore assessed by measuring the numbers of true and false positive (resp. $\# tp$ and $\# fp$), true and false negative ($\# tn$ and $\#fn$) predictions in the top $\theta$ pairs, and usual related quantities: 
precision $ \mathbf{Pr} = \frac{\#tp}{\#tp+\#fp}$, 
recall $ \mathbf{Rc} = \frac{\#tp}{\#tp+\#fn}$ 
and F1-score $ \mathbf{F} = \frac{2.\mathbf{Pr}.\mathbf{Rc}}{\mathbf{Pr} + \mathbf{Rc}}$.

Previous works have emphasized the dramatic effect of class imbalance (or skewness) on link prediction problems in social networks~\citep{lichtenwalter2010new,comar2011linkboost}.
The fact that the network is sparse and that there are many more pairs of nodes than links makes the prediction and its evaluation tricky.
The typical order of magnitude of the classes ratio for a social network made of $N$ nodes is indeed $ 1/N$.
It means that the number of predicted links is much lower than the number of candidate pairs, consequently the fall-out $\frac{\#fp}{\#fp+\#tn}$ is usually very small, making the ROC curve a potentially deceptive way of visualizing the performances, as discussed in~\cite{yang2015evaluating}.
For this reason and because we aim at improving both precision and recall over a large range, in the following we visualize the performances in the precision-recall space.

%
%


\section{Unsupervised rankings}
\label{sec:unsupervised}

\subsection{Ranking metrics}

In this work, we focus on structural features that assign to each pair of nodes a score based on topological information, then pairs are ranked according to this score.
Note that the metrics used may produce ties, in such case the relative order of tied items is decided randomly.
There is a large number of available metrics, which are designed to be correlated to the probability of existence of a link, see for example~\cite{zhou2009predicting,lu2011link,al2011survey}.
Moreover, we consider graphs which links may be weighted -- for example, weights may correspond to the number of interactions between two nodes.
When available, we draw benefit from the weight information to refine the prediction.
Nevertheless, the goal of this paper is neither to propose elaborate classifiers nor to delve deeply into the feature selection process, but to present a method that takes advantage of how complementary they are.
We have, therefore, chosen classic metrics and generalized them to the case of weighted networks -- other generalizations exist in the literature, e.g.~\cite{murata2007link}.

\subsubsection{Local features}

{\modif
In the following, $ \mathcal{N}(i)$ denotes the set of neighbors of node $i$, its degree is $ \delta(i) = | \mathcal{N}(i)|$, $ w(i,j)$ is the weight of a link $ (i,j) $ and $W(i)$ is the activity of a node $i$, that is the sum of the weights of its links.
Some metrics are local (also called \textit{neighborhood rankers}) as they only rank links among nodes which are at most at distance 2.
The common principle to their definition is that two nodes that have many neighbours in common are likely to be connected to each other.}

\begin{itemize}
\item \textit{Common Neighbors index (CN)}, based on the number of common neighbors shared by nodes $i$ and $j$, the corresponding unweigthed and weighted scores are
$$s_{CN}(i,j) = | \mathcal{N}(i) \cap \mathcal{N}(j) | 
\text{\hspace{3mm} and \hspace{3mm}} 
s_{CN_w}(i,j) = \sum \limits_{k \in \mathcal{N}(i) \cap \mathcal{N}(j)} w(i,k) \cdot w(j,k) $$ 
\item \textit{Adamic-Adar index (AA)}, which relies on the same principle as \textit{CN} and also promotes pairs which share low-degree neighbours,
$$
s_{AA}(i,j) = \sum \limits_{k \in \mathcal{N}(i) \cap \mathcal{N}(j)} \frac{1}{log(\delta(k))} 
\text{\hspace{3mm} and \hspace{3mm}}
s_{AA_w}(i,j) = \sum \limits_{k \in \mathcal{N}(i) \cap \mathcal{N}(j)} \frac{1}{log(W(k))} $$ 
\item \textit{Resource Allocation index (RA)}  is based on the same logic as $AA$ index, but gives a different weight to the degree of shared neighbours,
$$
s_{RA}(i,j) = \sum \limits_{k \in \mathcal{N}(i) \cap \mathcal{N}(j)} \frac{1}{\delta(k)} 
\text{\hspace{3mm} and \hspace{3mm}}
s_{RA_w}(i,j) = \sum \limits_{k \in \mathcal{N}(i) \cap \mathcal{N}(j)} \frac{1}{W(k)} $$ 
\item \textit{S\o rensen index (SR)},  which promotes pairs that have a large fraction of their neighbourhood in common,
$$
s_{SR}(i,j) =  \frac{2 \cdot| \mathcal{N}(i) \cap \mathcal{N}(j) | }{\delta(i) + \delta(j)}
\text{\hspace{3mm} and \hspace{3mm}}
s_{SR_w}(i,j) = \frac{ \sum \limits_{k \in \mathcal{N}(i) \cap \mathcal{N}(j)} w(i,k) + w(j,k) }{ W(i) + W(j) } $$ 

\end{itemize}

\subsubsection{Distance-based features}

Other features are distance-based, since they are calculated using the large-scale structure of the network, and allow for the ranking of distant pairs of nodes:
\begin{itemize}
\item \textit{Katz index (Katz)}, computed from the number of paths from node $i$ to node $j$ of length $l$, i.e. $ \nu _{ij}(l)$, according to the following expression
$$ s_{Katz}(i,j) =  \sum \limits_{l=1}^{\infty} \gamma ^l \nu_{ij}(l) $$
Here, $\gamma$ is an attenuation parameter. 
It must be lower than $1$ and small values favour short paths over long ones.
The relevance of \textit{Katz} index stems from the fact that two nodes which are connected by many paths are more likely to be linked.
Note that in the weighted case, the number of paths is computed as if links were multilinks.\\

\item \textit{Random Walk with Restart index (RWR)}, derived from the \textit{PageRank} algorithm, $s_{\text{\textit{RWR}}_w}(i,j)$ is defined as the probability that a random walker starting on node $i$, going from a node $k$ to a node $k'$ with probability $ p . w(k,k') / W(k)$ and returning to $i$ with probability $ 1 - p $, is on $j$ in the steady state of the process.
The fact that a random walk starting on $i$ has a high probability to go through $j$ indicates that both nodes are tightly related in the network, and may therefore be linked.\\
\item \textit{Preferential Attachment index (PA)}, based on the observation that active nodes tend to connect preferentially in social networks.
$$s_{\text{\textit{PA}}_w}(i,j) = W(i).W(j) $$
\end{itemize}

\subsubsection{Intermediary features}

In practice, the exact computation of distance-based metrics is expensive on large networks, that is why approximations are often favoured to compute these scores.
Both \textit{Katz} and \textit{RWR} are computed using infinite sums, from now on it is approximated by keeping only the first four dominating terms to reduce the computational cost\footnote{The limiting factor in the experiments presented in the following is the loading in memory of the adjacency matrix and its powers, which sizes are limited to around 1GB in our implementations.}, which is a usual practice for computing large-scale centrality estimators~\citep{lu2011link}.
This approximation means that we can only predict links between pairs of nodes at a maximum distance of 4.
Notice that it is a way to reduce the class-imbalance problem: as distant pairs are less likely to be connected, we dismiss them in order to increase the (true positive / candidate pairs) ratio.
As the class imbalance problem is known to hinder dramatically the performance of PA, we have restricted the ranking in this case to pairs of nodes at a maximum distance of~3.
Notice that with larger maximum distances, we can increase the maximum recall that can be reached, but at the cost of a drop of precision.

When even distance 4 approximation is too expensive, we use the \textit{Local Path} index, especially designed to capture the structure at an intermediary scale
$$ s_{\text{\textit{LP}}}(i,j) = \nu _{ij}(2) + \gamma \cdot \nu _{ij}(3) $$
With the same notations as \textit{Katz} index and $\gamma$ is the corresponding attenuation parameter.

\subsection{Borda's method}

The main purpose of this work is to develop a framework to exploit a set of $\alpha$ rankings for link prediction.
%
%
Here, we present an unsupervised way of merging rankings stemming from social choice theory: Borda's method is a \textit{rank-then-combine} method originally proposed to obtain a consensus from a voting system~\citep{deborda1781memoire}.
Each pair is given a score corresponding to the sum of the number of pairs ranked below, that is to say:

$$ s_B (i,j) = \sum \limits_{k = 1} ^{\alpha} | r_k | -  r_k (i,j)  $$
where $ | r_k |$ denotes the number of elements ranked in $ r_k $, and $r_k (i,j) $ the ranking of pair $ (i,j) $ in ranking $ r_k $.

This scoring system may be biased in the sense that it would favour some predictors - in our case distance-based scores - by the fact that they feature more elements. 
In other words, this definition does not cover well the partial ranking situation.
To alleviate this problem, we use the same method as the one cited in~\cite{dwork2001rank}: \textit{``by apportioning all the
excess scores equally among all unranked candidates''.
Another way of phrasing it is that all unranked pairs in ranking $ r_k $ are considered as ranked on an equal footing and below all ranked pairs.
Borda's method is computationally cheap, which is a highly desirable property in the case under consideration, where many items are ranked.
}


\section{\textit{RankMerging} framework}
\label{sec:supervised}

The ranking methods presented in the previous section use structural information in complementary ways.
In social networks, communication patterns are different among different groups, e.g. family, friends, coworkers etc.
Consequently, one expects that a link detected as likely by using a specific ranking method may not be discovered using another one.
In this section, we describe the supervised machine learning framework that we created to aggregate information from various ranking techniques for link prediction in social networks.
In a nutshell, it does not demand for a pair to be highly ranked according to all criteria (as in a consensus rule), but to be highly ranked in at least one.
The whole procedure is referred to as \textit{RankMerging}\footnote{Two implementations with user guides are available, one in Ocaml at \url{https://archive.softwareheritage.org/browse/origin/https://bitbucket.org/tabourier/rankmerging/src/master/directory/} and another in C++ at \url{https://gitlab.lip6.fr/tabourier/rankmergingcpp}}.

\subsection{Optimization task}
\label{sec:opti_task}

The learning phase of \textit{RankMerging} aggregation is based on a greedy optimization algorithm.
We first define the optimization task that we want to achieve during this phase.

{\modif
In the learning set, the items ranked are pairs of nodes which are labeled either with $1$ if they are connected and $0$ otherwise.
For a fixed number of predictions $\theta$, $\mathcal{S}_\theta$ is the sum of the values of the labels in the output ranking $r_o$, that is to say the number of true positive predictions among the top $\theta$ pairs in $r_o$.
$\mathcal{S}_\theta$ is a function of $ \Phi = [ r_{n_1}(k_1), \ldots , r_{n_\theta}(k_\theta)] $, that is the ordered list of length $ \theta $ which contains the sequence of pairs selected: $ r_{n_1}(k_1) $  means that the first element selected is the $ k_1^{th} $ pair of ranking $ r_{n_1}$, etc.
As an item can be present in several rankings but put only once in the output ranking, the items appended to the output are no longer available, in other words, all elements of the list $ \Phi $ have to be distinct.
If $ \phi_i $ denotes the number of items selected from ranking $r_i$, note that $\{ \phi_1 , \ldots , \phi_\alpha \}$ can be trivially derived from $ \Phi $ and that we have $\phi_1 + \ldots + \phi_\alpha = \theta$.
In the most general case, we are looking for the sequence $\Phi$ that yields the maximal value of $\mathcal{S}_\theta (\Phi)$.

Now, we precise a few choices in the design of our method, which simplify the optimization task described above.
The aggregation process is iterative, starting from an empty output ranking.
At each iteration, we select one of the input rankings and append its highest ranked item available to the output ranking.
Thus in our case, an element $r_{n_i}(k_i)$ can feature $ \Phi $ only if all elements above in $r_{n_i}$ are already in $ \Phi $.
In short, the aggregation is realized as a sequence of selections of input rankings.
In this context, the function $\mathcal{S}_\theta $ to optimize can be reformulated as a function  $\mathcal{S}_\theta (\hat{\Phi})$, where $ \hat{\Phi} = [ r_{n_1}, \ldots , r_{n_\theta}] $ is the ordered sequence of rankings selected during the process.
%
%
We are looking for a sequence which yields the maximum number of pairs labeled $1$ among the $\theta$ pairs in $ r_o $.
There is no obvious way to solve this problem exactly, therefore, we use a heuristic method to solve it by local search.
}

\subsection{Introductive example}

While the principle of our method is quite intuitive, its technical details may be dry.
That is why we describe in this section the process qualitatively, and develop it on an example to help the reader understanding the following sections.
To avoid confusion, we denote from now on the structures relative to the learning phase with index~$ L $ and the ones relative to the test phase with index~$ T $.

\subsubsection{Learning phase}

{\modif
During the learning phase, we have a learning graph which is used to build the prediction model.
The pairs of nodes which are not edges of this graph constitute the set of labeled pairs: some are not connected and others are edges of the calibration set, denoted $\mathcal{E}_{cal}$ in the following.
In practice, the prediction model is built to guess these links during the learning phase.}

Let us remind that the inputs of the learning algorithm are the unsupervised rankings, which are obtained using various kinds of scores (or rankers) on the learning graph.
Note that a ranker may give the same score to several pairs of nodes, in this case the order of the pairs is decided randomly.
The number of predictions is set to $\theta_L$.
The outputs of the learning phase are the merged ranking  $ r_{o,L} $ containing $\theta_L$ pairs and the values of the coefficients $ \{ \phi_1 , \ldots , \phi_\alpha \} $ defined in Section~\ref{sec:opti_task}.
These coefficients can be understood as the weights of the input rankings in $ r_{o,L} $.

To approximate the optimization task defined above, we proceed in the following way.
At each step, a ranking among the input rankings is selected and its highest available pair is appended to the output ranking $ r_{o,L} $.
The selection process is based on an estimate of the probability that the highest available pair is actually connected.
To do so, for each input ranking we build a sliding window $ \mathcal{W}_{i,L} $ which contains the next $g$ available pairs.
The window size $g$ is the only parameter of the method that the user has to set.
We denote $ \chi _i $ the quantity of true positive (or \textit{tp}) in window $ \mathcal{W}_{i,L} $.
Then, the average probability that any pair is connected in $ \mathcal{W}_{i,L} $ is the ratio $ \chi _i / g$.
We select the ranking which has the highest probability value, that is to say the highest $ \chi _i $ value.
Ties are broken randomly, which makes the method non-deterministic.

Let us put into practice this process on an example.
In Table~\ref{tab:algo}, we consider the learning process on two input rankings $ r_{1,L} $ and $ r_{2,L} $.
The window size $g$ is fixed to $5$ and the number of predictions $ \theta _L = 4$.
Initially, there are $4$ \textit{tp} in $\mathcal{W}_{1,L}$ and $3$ \textit{tp} in $\mathcal{W}_{2,L}$, consequently the first link selected is the top-ranked pair available in $\mathcal{W}_{1,L}$, which is $(1,2)$.
This pair is therefore appended to the output ranking $ r_{o,L} $ and excluded from the ranking $r_{2,L}$.
At step 2, we have $4$ \textit{tp} in both windows, the ranking is then selected randomly.
Let us suppose that $r_{2,L}$ has been selected, therefore, the next link added to $ r_{o,L} $ is $(5,18)$.
At step 3, $ \chi _1 = 4 $ and $ \chi _3 = 3 $ then the pair selected to join $ r_{o,L} $ is $(1,4)$ from $ r_{1,L} $.
At step 4, $ \chi _1 = 4 $ and $ \chi _2 = 3 $ and the pair selected is $(5,6)$ from $ r_{1,L} $.
At this point, $ \phi_1 = 3 $ and $ \phi_2 = 1 $.


\newcommand{\greencross}{{\scriptsize \cellcolor{lightgreen}{x}}}
\newcommand{\graycross}{{\scriptsize \cellcolor{gris}{x}}}
\newcommand{\redcross}{{\scriptsize \cellcolor{red}{x}}}
\newcommand{\whitecross}{{\scriptsize x}}

\begin{table}[!h]
\begin{footnotesize}
\setlength\fboxsep{1pt}
\newcolumntype{M}[1]{>{\centering}m{#1}}
\begin{center}
\begin{tabular}{|M{0.8cm}|c||M{0.8cm}|c|}
\hline
$ r_{1,L} $ &  $tp$ & $ r_{2,L} $ & $tp$  \\
\hline
\cellcolor{gris}{(1,2)} & \cellcolor{gris}{} & \cellcolor{gris}{(5,18)} & \cellcolor{gris}{x}\\
\cellcolor{gris}{(1,4)}  & \cellcolor{gris}{x} & \cellcolor{gris}{(1,2)}  & \cellcolor{gris}{}\\
\cellcolor{gris}{(5,6)}  & \cellcolor{gris}{x}  & \cellcolor{gris}{(8,9)} & \cellcolor{gris}{} \\
\cellcolor{gris}{(6,12)}  & \cellcolor{gris}{x} & \cellcolor{gris}{(5,6)} & \cellcolor{gris}{x} \\
\cellcolor{gris}{(5,18)} & \cellcolor{gris}{x} & \cellcolor{gris}{(7,11)} & \cellcolor{gris}{x} \\
(3,4) &  & (6,9) & x \\
(4,9) & x & (1,14) & \\
(7,11) & x & (2,9) &  \\
(2,9) &  & (3,7) &  \\
\hline
\end{tabular}
$ \rightarrow $
\begin{tabular}{|M{0.8cm}|c||M{0.8cm}|c|}
\hline
$ r_{1,L} $ &  $tp$ & $ r_{2,L} $ & $tp$  \\
\hline
\cellcolor{lightgreen}{(1,2)} & \cellcolor{lightgreen}{} & \cellcolor{gris}{(5,18)} & \cellcolor{gris}{x}\\
\cellcolor{gris}{(1,4)}  & \cellcolor{gris}{x} & \cellcolor{red}{\sout{(1,2)}}  & \cellcolor{red}{} \\
\cellcolor{gris}{(5,6)}  & \cellcolor{gris}{x}  & \cellcolor{gris}{(8,9)} & \cellcolor{gris}{} \\
\cellcolor{gris}{(6,12)}  & \cellcolor{gris}{x} & \cellcolor{gris}{(5,6)} & \cellcolor{gris}{x} \\
\cellcolor{gris}{(5,18)} & \cellcolor{gris}{x} & \cellcolor{gris}{(7,11)} & \cellcolor{gris}{x} \\
\cellcolor{gris}{(3,4)} & \cellcolor{gris}{} & \cellcolor{gris}{(6,9)} & \cellcolor{gris}{x} \\
(4,9) & x & (1,14) & \\
(7,11) & x & (2,9) &  \\
(2,9) &  & (3,7) &  \\
\hline
\end{tabular}
$ \rightarrow $
\end{center}
\end{footnotesize}
%

%
\begin{footnotesize}
\setlength\fboxsep{1pt}
\newcolumntype{M}[1]{>{\centering}m{#1}}
\begin{center}
\begin{tabular}{|M{0.8cm}|c||M{0.8cm}|c|}
\hline
$ r_{1,L} $ &  $tp$ & $ r_{2,L} $ & $tp$  \\
\hline
\cellcolor{lightgreen}{(1,2)} & \cellcolor{lightgreen}{} & \cellcolor{lightgreen}{(5,18)} & \cellcolor{lightgreen}{x}\\
\cellcolor{gris}{(1,4)}  & \cellcolor{gris}{x} & \cellcolor{red}{\sout{(1,2)}}  & \cellcolor{red}{} \\
\cellcolor{gris}{(5,6)}  & \cellcolor{gris}{x}  & \cellcolor{gris}{(8,9)} & \cellcolor{gris}{} \\
\cellcolor{gris}{(6,12)}  & \cellcolor{gris}{x} & \cellcolor{gris}{(5,6)} & \cellcolor{gris}{x} \\
\cellcolor{red}{\sout{(5,18)}}  & \cellcolor{red}{x} & \cellcolor{gris}{(7,11)} & \cellcolor{gris}{x} \\
\cellcolor{gris}{(3,4)} & \cellcolor{gris}{} & \cellcolor{gris}{(6,9)} & \cellcolor{gris}{x} \\
\cellcolor{gris}{(4,9)} & \cellcolor{gris}{x} & \cellcolor{gris}{(1,14)} & \cellcolor{gris}{}\\
(7,11) & x & (2,9) &  \\
(2,9) &  & (3,7) &  \\
\hline
\end{tabular}
%
%
$ \rightarrow $
\begin{tabular}{|M{0.8cm}|c||M{0.8cm}|c|}
\hline
$ r_{1,L} $ &  $tp$ & $ r_{2,L} $ & $tp$  \\
\hline
\cellcolor{lightgreen}{(1,2)} & \cellcolor{lightgreen}{} & \cellcolor{lightgreen}{(5,18)} & \cellcolor{lightgreen}{x}\\
\cellcolor{lightgreen}{(1,4)}  & \cellcolor{lightgreen}{x} & \cellcolor{red}{\sout{(1,2)}}  & \cellcolor{red}{} \\
\cellcolor{gris}{(5,6)}  & \cellcolor{gris}{x}  & \cellcolor{gris}{(8,9)} & \cellcolor{gris}{} \\
\cellcolor{gris}{(6,12)}  & \cellcolor{gris}{x} & \cellcolor{gris}{(5,6)} & \cellcolor{gris}{x} \\
\cellcolor{red}{\sout{(5,18)}}  & \cellcolor{red}{x} & \cellcolor{gris}{(7,11)} & \cellcolor{gris}{x} \\
\cellcolor{gris}{(3,4)} & \cellcolor{gris}{} & \cellcolor{gris}{(6,9)} & \cellcolor{gris}{x} \\
\cellcolor{gris}{(4,9)} & \cellcolor{gris}{x} & \cellcolor{gris}{(1,14)} & \cellcolor{gris}{}\\
\cellcolor{gris}{(7,11)} &  \cellcolor{gris}{x} & (2,9) &  \\
(2,9) &  & (3,7) &  \\
\hline
\end{tabular}
$ \rightarrow $
\end{center}
\end{footnotesize}
%

%
\begin{footnotesize}
\setlength\fboxsep{1pt}
\newcolumntype{M}[1]{>{\centering}m{#1}}
\begin{center}
\begin{tabular}{|M{0.8cm}|c||M{0.8cm}|c|}
\hline
$ r_{1,L} $ &  $tp$ & $ r_{2,L} $ & $tp$  \\
\hline
\cellcolor{lightgreen}{(1,2)} & \cellcolor{lightgreen}{} & \cellcolor{lightgreen}{(5,18)} & \cellcolor{lightgreen}{x}\\
\cellcolor{lightgreen}{(1,4)} & \cellcolor{lightgreen}{x} & \cellcolor{red}{\sout{(1,2)}}  & \cellcolor{red}{} \\
\cellcolor{lightgreen}{(5,6)}  & \cellcolor{lightgreen}{x}  & \cellcolor{gris}{(8,9)} & \cellcolor{gris}{} \\
\cellcolor{gris}{(6,12)}  & \cellcolor{gris}{x} & \cellcolor{red}{\sout{(5,6)}} & \cellcolor{red}{x} \\
\cellcolor{red}{\sout{(5,18)}}  & \cellcolor{red}{x} & \cellcolor{gris}{(7,11)} & \cellcolor{gris}{x} \\
\cellcolor{gris}{(3,4)} & \cellcolor{gris}{} & \cellcolor{gris}{(6,9)} & \cellcolor{gris}{x} \\
\cellcolor{gris}{(4,9)} & \cellcolor{gris}{x} & \cellcolor{gris}{(1,14)} & \cellcolor{gris}{}\\
\cellcolor{gris}{(7,11)} & \cellcolor{gris}{x} & \cellcolor{gris}{(2,9)} & \cellcolor{gris}{} \\
\cellcolor{gris}{(2,9)} & \cellcolor{gris}{} & (3,7) &  \\
\hline
\end{tabular}
\end{center}
\end{footnotesize}
\caption{\label{tab:algo}
Four steps illustrating the learning algorithm with rankings $ r_{1,L} $ and $ r_{2,L} $ ($g=5$, $\theta_L=4$).
Pairs predicted (i.e. $\in r_{o,L}$) have green backgrounds.
Pairs with gray backgrounds are in the windows $\mathcal{W}_1$ and $\mathcal{W}_2$. 
Barred pairs with red backgrounds have already been predicted and are no longer available.}
\end{table}

\subsubsection{Test phase}
\label{desc-test-phase}

The test phase of \textit{RankMerging} consists in using the coefficients $\{ \phi_1 , \ldots , \phi_\alpha \}$ computed at each step during the learning phase in order to select the ranking on the test set.

The inputs of the test phase are the unsupervised ranking defined on the test graph, denoted  $  \{ r_{1,T} , \ldots , r_{\alpha,T} \} $, and the $ \{ \phi_1 , \ldots , \phi_\alpha \} $  learned during the learning process.
Its output is the merged ranking of items $ r_{o,T} $, which is initially empty.
For the sake of simplicity, let us first suppose that there are the same number of items to rank in the learning phase and in the test phase.
In this case, we make the assumption that the optimal aggregated ranking in the test phase should be obtained by the same mixing as the aggregated ranking of the learning set.
For example, if the item ranked at position $k$ in $ r_{o,L} $ has been selected from ranking $ r_{i,L} $ during the learning process, we should rank at position $k$ in $ r_{o,T} $ a pair from ranking $ r_{i,T} $.

In Table~\ref{tab:test}, we give an example of test using the $ \phi_i $ learnt on the example of Table~\ref{tab:algo}.
At the first step of the learning algorithm, $ \phi_1 = 1 $ and $ \phi_2 = 0 $, so the top ranked item of $r_{1,T}$, which is $ (2,8) $, is selected at the first step of the testing process and appended to $ r_{o,T} $.
We iterate this process: at step 2,  $ \phi_1 = 1 $ and $ \phi_2 = 1 $, we therefore append the highest available item of $ r_{2,T} $ to $ r_{o,T} $, which is $ (1,8) $,  etc.

\begin{table}[!h]
\begin{footnotesize}
\setlength\fboxsep{1pt}
\begin{center}
\begin{tabular}{|c||c|c|}
\hline
learning step & $\phi_1 $  & $\phi_2 $  \\
\hline
1 & 1 & 0 \\
2 & 1 & 1 \\
3 & 2 & 1 \\
4 & 3 & 1 \\
\hline
\end{tabular}

\vspace{3mm}

\begin{tabular}{|c|c|}
\hline
$ r_{1,T} $ &  $ r_{2,T} $  \\
\hline
\cellcolor{lightgreen}{(2,8)} & (1,8) \\
(1,8) & (9,11) \\
(5,11) & (4,5) \\
(3,6) & (5,11) \\
\hline
\end{tabular}
$ \longrightarrow $
\begin{tabular}{|c|c|}
\hline
$ r_{1,T} $ &  $ r_{2,T} $ \\
\hline
\cellcolor{lightgreen}{(2,8)} & \cellcolor{lightgreen}{(1,8)} \\
\cellcolor{red}{\sout{(1,8)}} & (9,11) \\
(5,11) & (4,5) \\
(3,6) & (5,11) \\
\hline
\end{tabular}
$ \longrightarrow $
\begin{tabular}{|c|c|}
\hline
$  r_{1,T} $ &  $ r_{2,T} $ \\
\hline
\cellcolor{lightgreen}{(2,8)} & \cellcolor{lightgreen}{(1,8)} \\
\cellcolor{red}{\sout{(1,8)}} & (9,11) \\
\cellcolor{lightgreen}{(5,11)} & (4,5) \\
(3,6) & \cellcolor{red}{\sout{(5,11)}} \\
\hline
\end{tabular}
$ \longrightarrow $
\begin{tabular}{|c|c|}
\hline
$  r_{1,T} $ &  $ r_{2,T} $  \\
\hline
\cellcolor{lightgreen}{(2,8)} & \cellcolor{lightgreen}{(1,8)} \\
\cellcolor{red}{\sout{(1,8)}} &  (9,11)\\
\cellcolor{lightgreen}{(5,11)} & (4,5) \\
\cellcolor{lightgreen}{(3,6)} & \cellcolor{red}{\sout{(5,11)}} \\
\hline
\end{tabular}
\end{center}
\end{footnotesize}
\caption{\label{tab:test}
Illustration of the test algorithm.
Top: $ \phi_i $ learnt during the learning phase.
Bottom: pairs in $r_{o,T}$ at each step of the aggregation (green background), barred pairs with red backgrounds have already been predicted and cannot be selected anymore.
}
\end{table}

\begin{figure}

\begin{center}
\includegraphics[height=4.5cm]{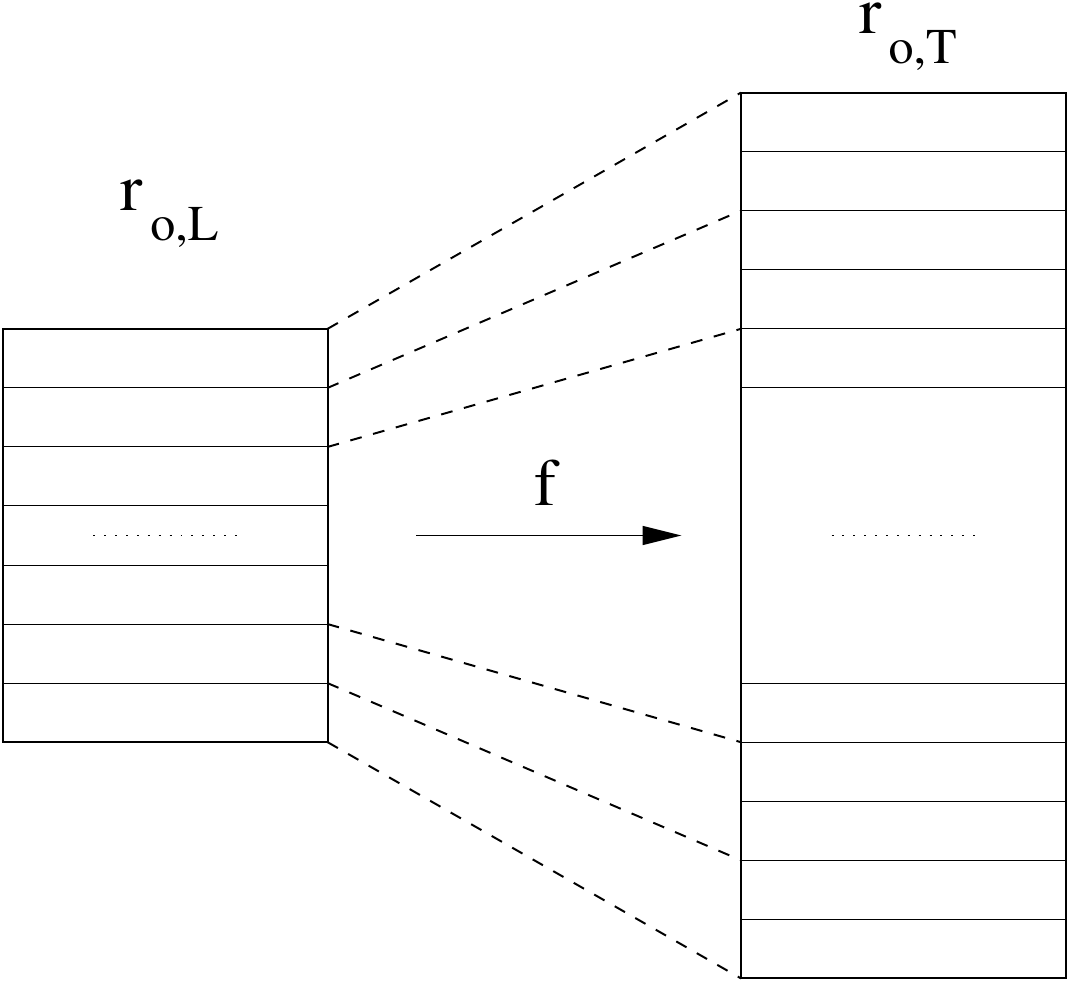}
\caption{\label{fig:scaling_factor} Schematic representation of the scaling factor $f$, ratio of the number of pairs ranked in the test set over the number of pairs ranked in the training set. On this example, the number of items ranked in the test set is supposed to be twice as large as the number of items ranked in the learning set, so that $f=2$. }

\end{center}

\end{figure}

It is possible that the number of items ranked using the learning graph and the number of items ranked using the test graph differ.
In this case, we define and compute a scaling factor $f$, which is the ratio between the number of items ranked on the test set over the number of items ranked on the learning set.
Then, when predicting item in position $k_T$ in the test output ranking, it should correspond to the item ranked in position $ k_L = \lfloor k_T / f \rfloor $. 
This idea can be intuitively understood using Figure~\ref{fig:scaling_factor}, the mixing should be proportional to the size of the ranking considered.
For example, if there are twice more pairs ranked in the testing set than in the learning set, the ranking $r_{i,T}$ selected for $k_T=1$ and $k_T=2$ corresponds to the ranking $r_{i,L}$ selected for $k_L=1$.


\subsection{Algorithms}
\label{sec:principle}

In this section, we describe the algorithms of \textit{RankMerging} aggregation process with more precise technical details.

\subsubsection{Learning phase}
\label{subsec:learning}

The learning set in the context of link prediction is composed of a learning graph $ \mathcal{G}_{L} = \{ \mathcal{V}_{L} , \mathcal{E}_{L} \} $, where $\mathcal{V}_{L}$ is the set of vertices and $\mathcal{E}_{L}$ is the set of edges.
Edges of $ \mathcal{E}_{L} $ are not ranked in the process and cannot be predicted.
For this purpose we also have an additional set of links $\mathcal{E}_{cal}$ among nodes of $ \mathcal{V}_{L} $.
During the learning phase, we aim at predicting these additional edges, which are the pairs labeled as true positive as already mentioned.

We have $ \alpha $ different rankings $ \{ r_{1,L}, ... , r_{\alpha,L} \} $ obtained with unsupervised ranking methods on $ \mathcal{G}_{L}$.
Any ranking $ r_{i,L} $ contains exclusively pairs of $ (\mathcal{V}_{L} \times \mathcal{V}_{L}) \setminus \mathcal{E}_{L} $.
This set of rankings is described as a two dimensional table $ \mathcal{R}_L $, so that $ \mathcal{R}_L[i] = r_{i,L} $ and $ \mathcal{R}_L[i][k] $ denotes the pair ranked in position $k$ of ranking $ r_{i,L} $.

We previously explained that each ranking is equipped with a sliding window $ \mathcal{W}_{i} $, that we denote $ \mathcal{W}[i] $ from now on, as it is practically stored as a table of sets.   
Each window contains exactly $g$ items, which are the highest ranked pairs in this ranking, still available for selection.
A window can be located using a start index $ \rho $ which is the rank of the highest available pair and an end index $ \sigma $ which is the rank of the highest pair out the window.
Storing these indices in tables, $ \rho [i] $ and $ \sigma [i] $ are respectively the start and end index of $ \mathcal{W}[i] $.
Notice that initially $ \sigma [i] - \rho[i] = g$, however it is not true during the whole process.

We also need to store the output ranking $ r_{o,L} $ in a table of length $ \theta _L $ (the number of predictions, which set by the user).  
Most importantly, we need to register the values of $ \{ \phi_1, \ldots , \phi_\alpha \} $.
We store them in a two dimensional table $ \Phi $, so that $ \Phi[i][k] $ is the value of $ \phi_i $ at the $k^{th}$ step of the aggregation process, that is the selection of item  $ r_{o,L}[k] $. 
Note also that the values of $ \{ \phi_1, \ldots , \phi_\alpha \} $ at the $k^{th}$ step of the aggregation are the values corresponding to $ \theta _L = k$.
So, when the number of predictions is $ \theta _L $, we can actually make a prediction for any value $ k \leq \theta _L $.

%

Now that all the useful structures are defined, we describe in a few words the aggregation process.
At any step $k$, we evaluate for each window the number of links that it contains, which is $ \chi_i = | \mathcal{W}[i] \cap \mathcal{E}_{cal} |$.
The ranking selected is the one for which this quantity is maximum, ties being broken randomly.
%
%
The highest ranked pair of the selected ranking is appended in position $k$ of the output ranking $ r_{o,L} $.
We register which ranking $ \mathcal{R}_L[j] $ has been selected by incrementing the value of $ \Phi[j] $, that is to say $ \Phi[j][k] = \Phi[j][k-1] + 1$, while for the other rankings ($i \neq j$), $ \Phi[i][k] = \Phi[i][k-1] $. 
Then, the windows $ \mathcal{W}[i] $ are updated so that each window contains exactly the $g$ highest ranked available pairs in $ \mathcal{R}_L[i] $.
The process is iterated until $ r_{o,L} $ contains $ \theta_L$ items.

The algorithm corresponding to this process is described in Algo.~\ref{algo:merging}.
In the description of the algorithm, we dropped the index $L$ for the sake of readability.

\begin{algorithm}[H]
\SetAlFnt{\footnotesize\sf}
\SetKwInOut{Input}{inputs}
\SetKwInOut{Output}{outputs}
\SetAlgoProcName{Nom}{nom}
\Input{table of rankings $\mathcal{R}$; real edge set $\mathcal{E}_{cal}$;\\
maximum number of predictions $\theta$; window size $g$;}
\Output{table $ \Phi $; merged ranking $r_o$;}
\tcp{initialization, $\forall i \in \{1 , \ldots , \alpha \}$:}
\Begin{
$\mathcal{W}[i] \leftarrow g$ first links in $\mathcal{R}[i]$\tcp*{pairs in window  $\mathcal{W}_i$}
$\chi[i] \leftarrow | \mathcal{W}[i] \cap \mathcal{E}_{cal} | $\tcp*{number of actual links in window $\mathcal{W}_i$} 
$\rho[i] \leftarrow 1 $\tcp*{start index of window $\mathcal{W}_i$}
$\sigma[i] \leftarrow g+1 $\tcp*{end index of window $\mathcal{W}_i$}
$k  \leftarrow 1$\tcp*{counter of the number of items ranked in $r_o$}
}
\While{ $k \leq \theta$}{
$ i_{max} \leftarrow $ index corresponding to maximum $ \chi[i] $;\\
$ r_o[k] \leftarrow  \mathcal{R}[i_{max}][\rho[i_{max}]]$;\\
$ \rho[i_{max}] \leftarrow \rho[i_{max}] + 1$;\\
$ k \leftarrow k+1$;\\
$ \Phi[i_{max}][k] \leftarrow \Phi[i_{max}][k-1] +1 $\tcp*{record $ \phi_{i_{max}}$ at step $k$}
$ \forall i \in \{1 , \ldots , \alpha \}$ and  $i \neq i_{max} $, $ \Phi[i][k] \leftarrow \Phi[i][k-1] $\tcp*{record other $ \phi_i$}
$ \forall i \in \{1 , \ldots , \alpha \}$, \textbf{update}$(\mathcal{W}[i],\chi[i],\rho[i],\sigma[i],r_o[k])$\tcp*{update windows $\mathcal{W}_i$}
}
\hrule
\vspace{1mm}
\SetKwInOut{Input}{Procedure}
\Input{\textbf{update}$(\mathcal{W}[i],\chi[i],\rho[i],\sigma[i],r_o[k])$:\\}
\SetAlgoProcName{update}{anautorefname}
\Begin{
\tcp{take away the selected pair from $ \mathcal{W}_i$:}
\If{$ r_o[k] \in \mathcal{W}[i] $}{
$ \mathcal{W}[i] \leftarrow \mathcal{W}[i] \setminus \{ r_o[k] \}$
}
\tcp{increase $\rho _i$ until reaching an available pair in $ r_i$:}
\While{ $ \mathcal{R}[i][\rho[i]] \in r_o $}{
	$ \rho[i] \leftarrow \rho[i] + 1$;}
\tcp{increase $\sigma _i$ until having $g$ pairs in $ \mathcal{W}_i$:}
\While{ $ | \mathcal{W}[i] | \leq g  $}{
$ l \leftarrow  \mathcal{R}[i][\sigma[i]]$;\\
$ \sigma[i] \leftarrow \sigma[i] + 1$;\\
\If{$ l \notin r_o $}{
$\mathcal{W}[i] \leftarrow \mathcal{W}[i] \cup \{ l \}$;
}
}
$\chi[i] \leftarrow | \mathcal{W}[i] \cap \mathcal{E}_{cal} | $\tcp*{update number of links in window $\mathcal{W}_i$}
}
\caption{\label{algo:merging} 
\textit{RankMerging} method: learning algorithm.}

\end{algorithm}

\subsubsection{Test phase~\protect\footnote{This subsection has been modified compared to the version of this paper published~\citep{tabourier2019rankmerging} to make the algorithm and the code more consistent.} 
}

The test set is composed of a test graph $ \mathcal{G}_{T} = \{ \mathcal{V}_{T} , \mathcal{E}_{T} \} $, where $\mathcal{V}_{T}$ is the set of vertices and $\mathcal{E}_{T}$ is the set of edges.
Edges of $ \mathcal{E}_{T} $ are not ranked in the process and cannot be predicted.
We aim at predicting edges of an additional set $\mathcal{E}_{perf}$, which is used to measure the performances of the method.

We have $ \alpha $ different rankings $ \{ r_{1,T}, ... , r_{\alpha,T} \} $ obtained with unsupervised ranking methods on $ \mathcal{G}_{T}$.
The feature used to obtain $ r_{i,T} $ on the test graph is the same as the feature used to obtain $ r_{i,L} $ on the learning graph.
Any ranking $ r_{i,T} $ contains exclusively pairs of $ (\mathcal{V}_{T} \times \mathcal{V}_{T}) \setminus \mathcal{E}_{T} $.
Here again, this set of rankings is described as a two dimensional table $ \mathcal{R}_T $, so that $ \mathcal{R}_T[i] = r_{i,T} $ and $ \mathcal{R}_T[i][k] $ denotes the pair ranked in position $k$ of ranking $ r_{i,T} $.

The test phase consists in aggregating rankings of $ \mathcal{R}_T $ according to the $ \{ \phi_1 , ... , \phi_\alpha \}$ learnt on the training network $ \mathcal{G}_{L} $.
This process yields an aggregated ranking $ r_{o,T} $, which is the final output of the method.
The number of pairs selected from any ranking $ \mathcal{R}_T[i] $ is stored in a table $ \Lambda $, such that $ \Lambda[i] $ is the number of pairs selected from $ \mathcal{R}_T[i] $.
The number of pairs selected from a ranking $ \mathcal{R}_T[i] $ denoted $ \Lambda[i] $ should be the value $\Phi[i][k]$ learned during the learning phase, rescaled by the scaling factor $f$, considering that the aggregated ranking of the test phase should have $f$ times the size of its equivalent during the learning phase.
%
%
Practically speaking, it means that for all $i \in \{ 1, \ldots , \alpha \}$, we append available items from ranking $ \mathcal{R}_T[i] $ to $ r_{o,T} $ until we have $ \Lambda[i] = \lfloor \Phi[i][k] / f \rfloor $.
Notice also that the size $ \theta_T $ of $ r_{o,T} $ cannot exceed $f$ times the size of $ r_{o,L} $, that is $ f \cdot \theta_L$.

%
%
%

Finally, the pairs appended to the output ranking are the highest ranked available pairs of $ \mathcal{R}_T[i] $.
For this purpose, we define a table $\mathcal{C}$, in which we store the location of the top-ranked available pair in $ \mathcal{C}[i] $.
The corresponding process is described in Algorithm~\ref{algo:testing}.
We dropped the index $T$ in the algorithm for the sake of readability.

%

\begin{algorithm}[!h]
\SetAlFnt{\footnotesize\sf}
\SetKwInOut{Input}{inputs}
\SetKwInOut{Output}{outputs}
\SetAlgoProcName{Nom}{nom}
\Input{table of rankings $\mathcal{R}$; $ \Phi $ table;\\
maximum number of predictions $\theta$; scaling factor $f$; }
\Output{aggregated ranking $r_{o}$;}
\tcp{initialization, $\forall i \in \{1 , \ldots , \alpha \}$:}
\Begin{
$\mathcal{C}[i] \leftarrow 1$\tcp*{location highest ranked available item in $r_i$}
$\Lambda[i] \leftarrow 0$\tcp*{number of items selected from $r_i$}
$k \leftarrow 1$\tcp*{index of the learning tables $ \phi _i$}
$k' \leftarrow 1$\tcp*{number of items ranked in $r_o$}
}
\While{ $k' \leq \theta $}{
\For{$ i \in \{1 , \ldots , \alpha \} $}{
  \While{$ \Lambda[i]  < \lfloor \Phi[i][k] / f \rfloor $ \textbf{and} $k' \leq \theta $ \textbf{and} $k \leq \vert \Phi[i] \vert$}{
	\If{$ \mathcal{R}[i][\mathcal{C}[i]] \notin r_o $}{
		$ r_o[k'] \leftarrow \mathcal{R}[i][\mathcal{C}[i]]$;\\
		$ \Lambda[i] \leftarrow \Lambda[i] +1$;\\
		$ k' \leftarrow k'+1$;\\
	}
	$\mathcal{C}[i] \leftarrow \mathcal{C}[i]+1$;\\	
	}
  }
  $ k \leftarrow k+1$
}
\caption{\label{algo:testing} 
\textit{RankMerging} method: testing algorithm.}

\end{algorithm}

\subsubsection{Complexity}

A substantial benefit of the learning algorithm is that we need to go through each learning ranking only once.
Moreover, by using appropriate data structures to store the windows $ \mathcal{W}_{i}$ -- e.g., associative arrays of sets -- it is easy to manage each update in $ O(1) $. 
So if we have $\alpha$ rankings and $ \theta_L $ predictions, it implies a $ O(\alpha \cdot \theta_L) $ temporal complexity.
Similarly, the test phase demands to go through the test rankings once, yielding a $ O(\alpha \cdot \theta_T) $ complexity.
Moreover, most of the memory used is due to the storage of the rankings, which is also $ O(\alpha \cdot \theta_L) $ or $ O(\alpha \cdot \theta_T) $.
These time and space consumptions are in general insignificant with regard to the complexities of the preliminary unsupervised classification methods, as is confirmed by the experimental computation times (see Section~\ref{exp-running-times}).

\subsection{Properties related to social choice theory}

In the context of social choice theory, ranking aggregation methods are often built in order to satisfy a set of properties that guarantee their goodness~\citep{arrow2012social}.
Two of the properties which are often considered as desirable are \textit{a)} satisfying the Condorcet criterion (in its extended version), and \textit{b)} being Kemeny optimal.
The former means that if a majority of rankings prefers an item $i$ to an item $j$ then it should be the case of the aggregated ranking.
The later means that the Kendall tau distance from the output ranking to the inputs is minimal.
Our method does not satisfy these two properties\footnote{It can also be shown that \textit{RankMerging} does not verify independence of irrelevant alternatives, weak monotonicity and continuity.
On the other hand, it satisfies the properties of consistency, unanimity, non-dictatoriality, neutrality, anonymity and non-constancy.}.

However, it is important to underline that \textit{RankMerging} is not built for the purpose of social choice.
These characteristics are relevant in an unsupervised context, where there is no ground truth to evaluate the result.
The underlying assumption is that each ranking should have the same weight and the distance to the input is the criterion to evaluate the quality of an aggregation.
As argued in~\cite{subbian2011supervised} too, there is a ground truth available in the context of supervised ranking aggregation tasks, that is the number of true positive predictions on the learning set.
As a consequence, satisfying these properties is not a priority for our purpose and  we naturally favoured the number of true predictions for the design of \textit{RankMerging}.
%
%
%
%


{\modif
\subsection{Design justification}

\label{sec:opt_justif}

Let us recall that the aggregation process aims at finding the maximum value of a function $ \mathcal{S}_\theta (\hat{\Phi})$, which is the sum of the labels of the output rankings of the learning phase.
The labels are either 1 (if the link exists) or 0 (if it does not) -- see Sec.~\ref{sec:opti_task}.
In this section, we explain why the design that we have chosen for the algorithm should be efficient to achieve this task, and what is the influence of parameter $g$ in the process.

During the learning phase of the algorithm, we select elements of the input rankings according to these rankings order.
This choice is efficient provided that input rankers are made so that highest ranked pairs have a higher probability of being true positive predictions -- which is expected from a \textit{good} classifier.
Moreover, we choose a ranking depending on a quality evaluator, that is $ \chi_i / g$.
Considering that $ \chi_i $ is the number of elements with label 1 in the sliding window of ranking $ r_{i,L} $, $ \chi_i / g$ can be understood as the probability that a random element in the window  is a true positive.

To investigate how efficient the aggregation process on the learning graph should be, we measure if the quality evaluator is indeed decreasing throughout the process.
We show on Figure~\ref{fig:check} a typical example of the evolution of $\chi_i / g$ for a given ranker (\textit{Common Neighbors}) in a given experiment (on the PSP dataset, see~\ref{subsec:psp-dataset}) and different $g$ values.

%


%
%

\begin{figure}[!h]
\begin{center}
\includegraphics[angle=-90,width=0.6\linewidth]{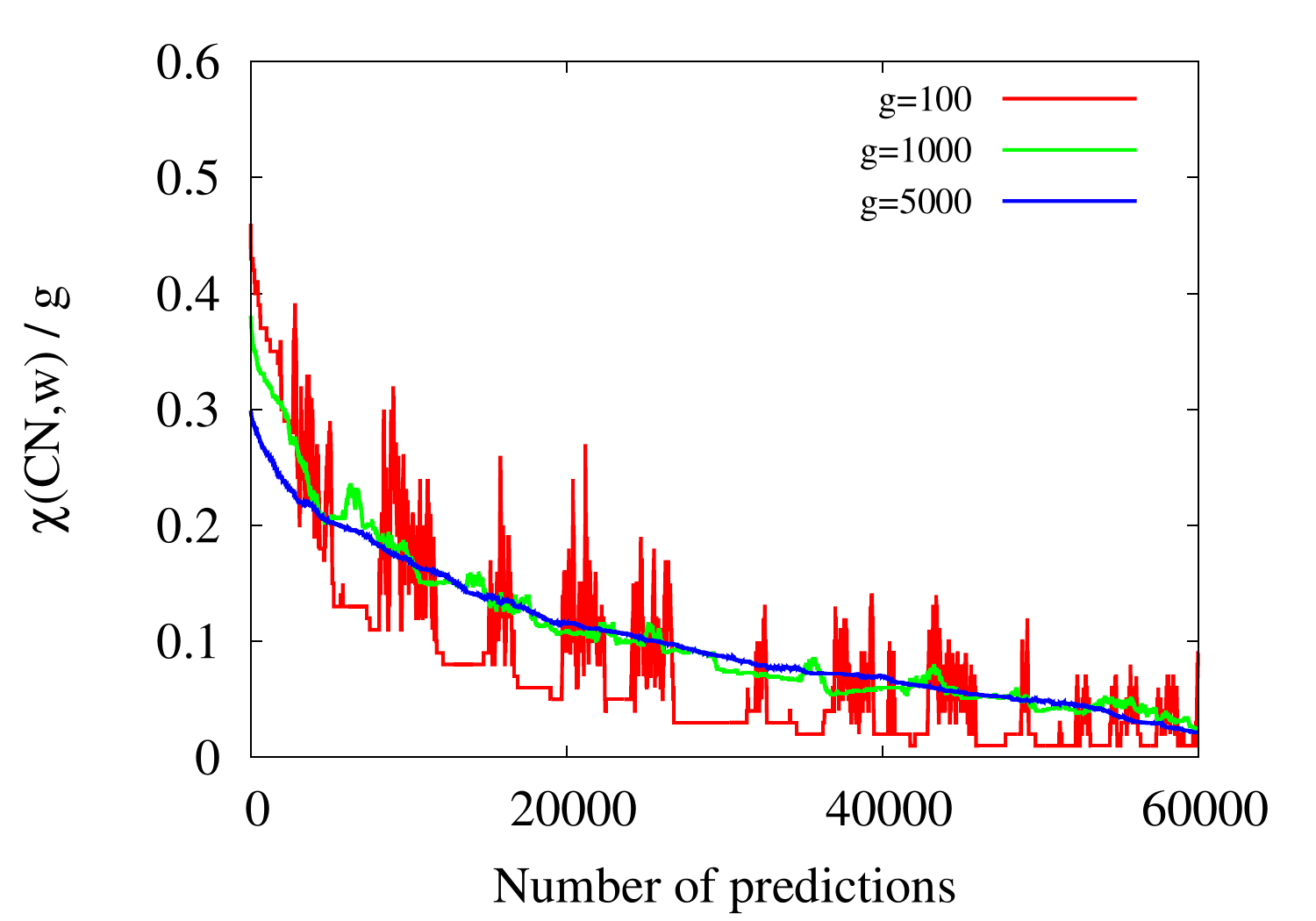}
\caption{\label{fig:check} 
Function $\frac{\chi_i}{g}$ corresponding to ranking $ \textit{CN}_w $ for $g=100,1000,5000$ during a learning process on the PSP dataset.
}
\end{center}
\end{figure}

We can observe that the decreasing condition is nearly fulfilled during the process if $g$ is large enough.
However, it does not mean that the best solution for the merging  problem is to take $g$ as large as possible,
as the larger $g$ gets, the lesser $ \chi_i / g$ is an accurate estimation of the probability that the next pair selected is an element with label 1.
Therefore, we have to manage a trade-off by tuning parameter $g$ in order to obtain the best possible performance of \textit{RankMerging}.
}


\section{Experiments}

\label{sec:exp}


\subsection{Datasets}

\label{sec:data}

\subsubsection{PSP phonecall network}

\label{subsec:psp-dataset}

We investigate a call detail record of approximately $ 14 \cdot 10^6 $ phonecalls of anonymized subscribers of a European phone service provider (PSP) during a one month period.
Users are modeled as nodes, and a link represents the existence of a communication relationship between two users.
A call between two users is a directed interaction, and we filter out calls which are not indicative of a lasting relationship (e.g., commercial calls) by considering only calls on bidirectional links, i.e. links that have been activated in both directions.
%
%
After this filtering has been applied, interactions between users are considered as undirected.
The total number of phone calls between nodes $i$ and $j$ is the weight of this link.
The resulting network is composed of 1,131,049 nodes, 795,865 links and 10,934,277 calls.

\subsubsection{DBLP coauthorship network}

The second dataset under study is the collection of computer science publications DBLP\footnote{Available at \url{http://konect.uni-koblenz.de/networks/dblp_coauthor}.}, which roughly contains  $ 19 \cdot 10^6 $ papers.
We study the collaboration network, where nodes of the network are authors and links connect authors who published together and links are weighted according to their number of collaborations.
The network features 10,724,828 links and 1,314,050 nodes.

\subsubsection{Pokec social network}

Pokec is a popular online social network in Slovakia\footnote{Available at \url{http://snap.stanford.edu/data/soc-pokec.html}.}.
We consider the network of friendship links between accounts.
In its original version there are around $ 31 \cdot 10^6 $ friendship links, but friendship is directed in Pokec, so we only kept reciprocated friendships in the spirit of the preprocessing of the PSP dataset. 
The final network contains 8,320,600 links and 1,632,804 nodes.


{\modif
\subsubsection{Facebook social network}

We also study a subpart of Facebook network, which has been described in~\cite{viswanath-2009-activity}\footnote{Available at \url{http://konect.uni-koblenz.de/networks/facebook-wosn-wall}.}.
This dataset is considered in a link prediction setting (rather than missing link recovery).
It registers communication among users via their Facebook wall from approximately  2005 to 2008.
We restricted ourselves to 700 days of data from January $1^{st}$ 2007 to November $ 30^{th} $ 2008, as the activity over the first two years is low.
Temporal labels are rounded to have a daily granularity.
The final temporal network contains 45,612 nodes, 175,651 links and 532,318 interactions (wall posts), links are considered undirected. 
}

\subsection{Benchmarks for comparison}

\label{subsec:bench}

In order to assess the efficiency of \textit{RankMerging}, we compare its performances to existing techniques.
\begin{itemize}
\item[$ \bullet $] We first compare to the various unsupervised rankings described in Section~\ref{sec:unsupervised}.
\item[$ \bullet $] We also compare to classic supervised techniques for classification tasks.
As stated in~\cite{al2006link}, there are a lot of such methods available, which performances are usually comparable.
Because of the number of items to be classified, we restricted ourselves to several computationally efficient methods, namely $k$ nearest neighbors (\textit{NN}), decision trees (also known as classification trees \textit{CT}) and \textit{AdaBoost} (\textit{AB})\footnote{We used a version of \textit{AdaBoost} which combines the results of several pruned classification trees.}.
The learning features used for the classification task are the unsupervised ranks of the pairs.
We have used implementations from Python scikit learn toolkit\footnote{\url{http://scikit-learn.org/} , for more details, see~\cite{scikit-learn}.}.
Note that these techniques are not specifically designed for link prediction tasks.
As a consequence, there is no obvious way of controlling the number of predictions.
We obtain several points in the precision-recall space by varying the algorithms parameters, respectively the number of neighbors $k$  with \textit{NN}, the minimum size of a leaf with \textit{CT}, and the number of pruned trees with \textit{AB}.
\item[$ \bullet $]
Finally, we compare our results to a supervised learning-to-rank technique.
We underline that most of the classic techniques are not available here, as the rankings considered have so many items that it not conceivable to use a method which complexity is worse than linear or linearithmic.
For example, methods based on the pairwise comparison of rankings, are not available here.
We use the \textit{weighted Borda} method proposed in~\cite{pujari2012supervised}, which follows the same principle as the unsupervised Borda method, except for the fact that the input rankings are weighted according to their level of performance on a training set.
{\modif
Several weighting schemes are proposed in that article, and we use the \textit{maximization of precision}, in which weights are proportional to the precision of the predictions related to the input rankings.}
We do not use the local Kemeny and supervised Kemeny methods either~\citep{subbian2011supervised}, because even if the merging process is presented as a linearithmic function of the size of the rankings, the whole procedure is actually quadratic as there is a preliminary pairwise score computation.
%
%
\end{itemize}

\subsection{Series 1}

The first series of experiments aims above all to describe how the method works on a practical example.
We focus on the PSP dataset and explore the impact of the parameter $g$ value and the question of the feature selection.

A possible motivation for recovering missing links in the context of a phone service provider is that PSPs only have access to records involving their own clients, so that they have an incomplete view of the social network as a whole.
In several practical applications, however, this missing information is crucial.
An important example is churn prediction, that is the detection of clients at risk of leaving to another PSP.
This risk is known to depend on the local structure of the social network~\citep{dasgupta2008social,ngonmang2012churn}.
Therefore, the knowledge of connections between subscribers of other providers is important for the design of efficient customer retention campaigns.
This series of experiments has been designed for this application, that is to say the prediction of links among the users of a PSP competitor.

\subsubsection{Protocol}

A Phone Service Provider is usually confronted to the following situation: it has full access to the details of phone calls between its subscribers, as well as between one of its subscriber and a subscriber of another PSP.
However, connections between subscribers of other PSPs are hidden.
%
%
{\modif In order to simulate this situation from our dataset, we divide the set of nodes $ \mathcal{V} $ into three disjoint sets: $\mathcal{V}_{1}$,~$\mathcal{V}_{2}$~and~$\mathcal{V}_{3}$.
$ \mathcal{V}_{1} \cup \mathcal{V}_{2}$ would be subscribers to the PSP, $ \mathcal{V}_{3} $ would be subscribers to competitors.
$\mathcal{V}_{1}$, $\mathcal{V}_{2}$ and $\mathcal{V}_{3}$ form a partition of $ \mathcal{V} $ and a partition of the set of links $ \mathcal{E} $ ensues.}
During the learning phase, links $\mathcal{V}_{1}$--$\mathcal{V}_{1}$ and $\mathcal{V}_{1}$--$\mathcal{V}_{2}$ are known, defining the set of links $\mathcal{E}_{L}$ of the graph $\mathcal{G}_{L}$, and we calibrate the model by guessing links  $\mathcal{V}_{2}$--$\mathcal{V}_{2}$, defining the set $\mathcal{E}_{cal}$. 
During the test phase, all the links are known except for links $\mathcal{V}_{3}$--$\mathcal{V}_{3}$, and we aim at guessing these links to evaluate the performances, we denote this set of links $\mathcal{E}_{perf}$.
Users have been assigned randomly to $\mathcal{V}_1$, $\mathcal{V}_2$ and $\mathcal{V}_3$ according to the proportions 50, 25, 25\%.
With these notations, we have:
\begin{itemize}
\item $ \mathcal{G}_{L} = ( \mathcal{V}_{L} = \mathcal{V}_{1} \cup \mathcal{V}_{2} \ , \ \mathcal{E}_{L} )$, it contains 848,911 nodes and 597,538 links and we aim at predicting the 49,731 links in $\mathcal{E}_{cal}$.
\item $ \mathcal{G}_{T} = ( \mathcal{V}_{T} = \mathcal{V} \ , \ \mathcal{E}_{T} = \mathcal{E} \setminus \mathcal{E}_{perf} )$, it contains 1,131,049 nodes and 746,202 links and we aim at predicting the 49,663 links in $\mathcal{E}_{perf}$.
\end{itemize}

%

\subsubsection{Unsupervised learning process}

\begin{figure}[h]
\begin{center}
\includegraphics[angle=-90,width=0.495\linewidth]{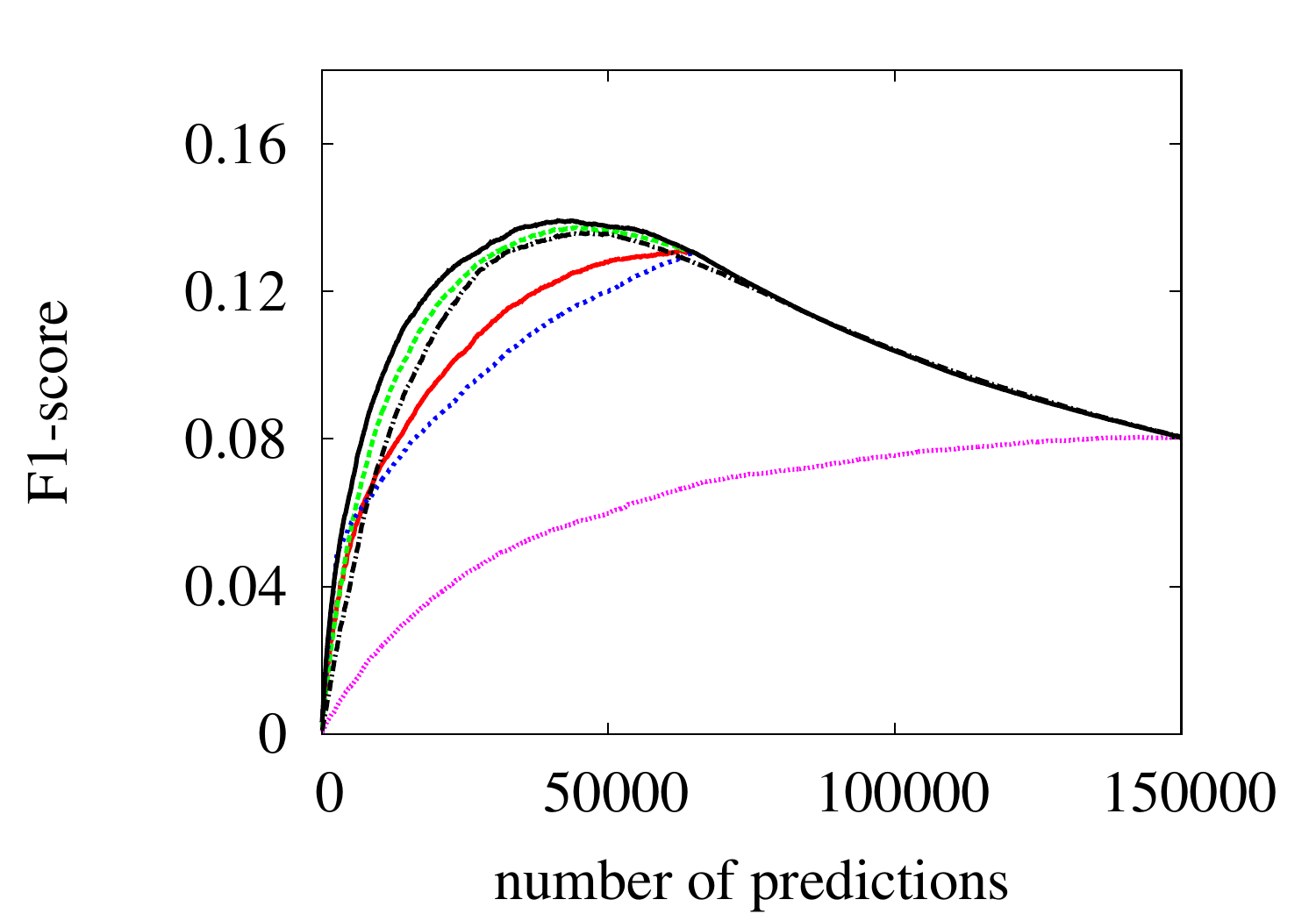}
\includegraphics[angle=-90,width=0.495\linewidth]{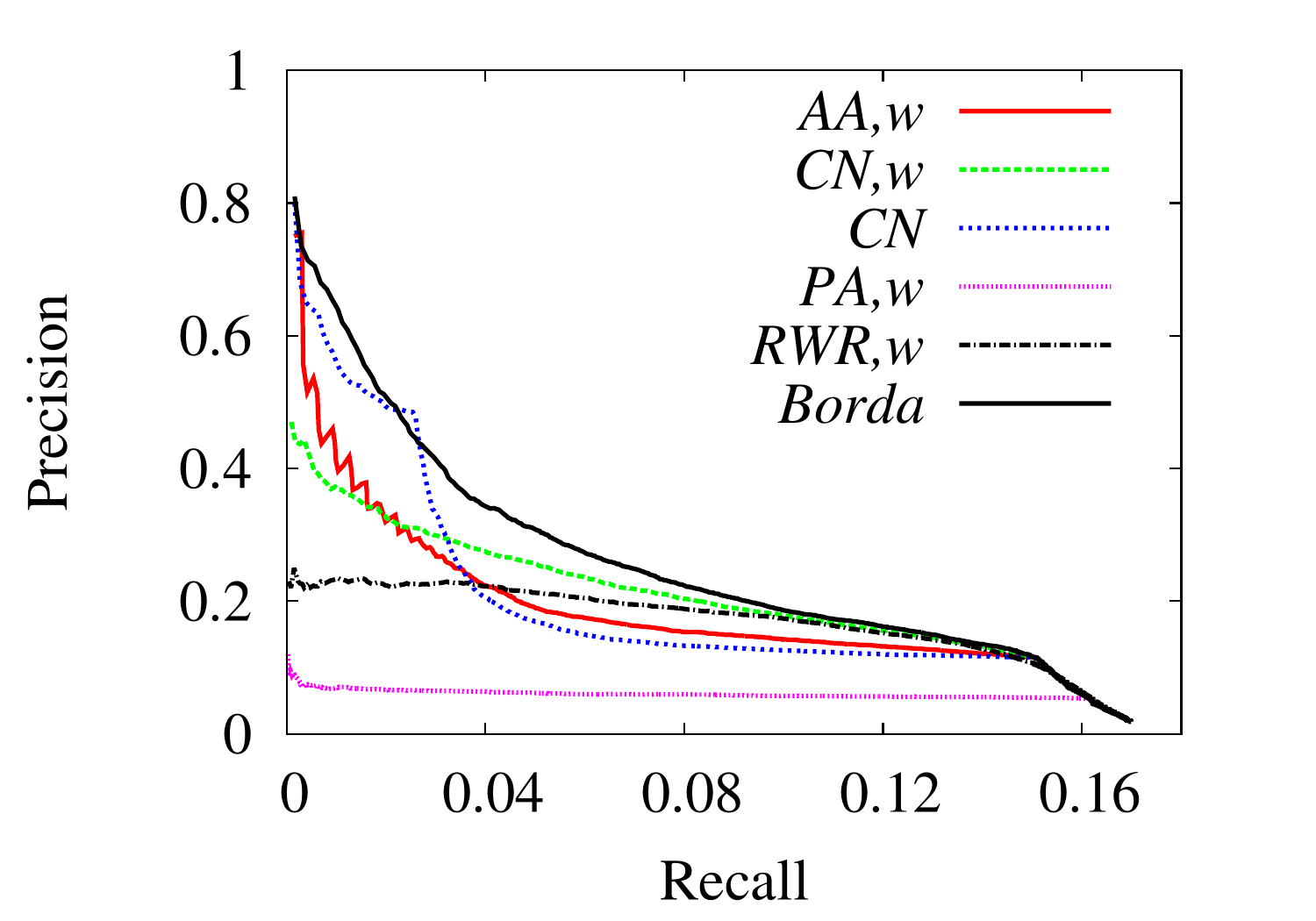}
\end{center}
\caption{\label{fig:unsup}
Results obtained on the learning set for various structural classifiers.
Left: F1-score as a function of the number of predictions. 
Right: precision versus recall curves.}
\end{figure}

We plot the results obtained on $ \mathcal{G}_{L} $ to predict $ \mathcal{E}_{cal} $ links for the above classifiers.
For the sake of readability, we only represent a selection of them in Figure~\ref{fig:unsup}.
The evolution of the F1-score significantly varies from one classifier to another.
For example, it increases sharply for \textit{CN}, and then more slowly until reaching its maximum, while \textit{RWR}$_w$ rises smoothly before slowly declining.
%
%
Borda's aggregation improves the performance of the classification, especially considering the precision on the top-ranked pairs.
\textit{RankMerging} method aims at exploiting the differences between these profiles.
Given the difficulty of the task, precision is low on average.
For instance, when recall is greater than $0.06$, precision is lower than $0.3$ for all rankers.
{\modif
We only used structural features here, making it impossible to predict links between nodes which are too distant from each other (further than four steps in our case).
On the one hand we are thus limited to small recall values.
%
%
But on the other hand, limiting predictions to pairs of nodes which are close is a way of reducing the class imbalance effect aforementioned.
Indeed, we dramatically reduce the set of candidate pairs, while the set of connected pairs is reduced by a much smaller factor as closer nodes have a higher probability of being connected~\citep{lichtenwalter2010new}.}
{\modif Finally, note that the process considered here is a partial aggregation process, as different rankings feature a different number of pairs (e.g., \textit{AA$_w$} and \textit{RWR$_w$}), but this does not affect its functioning.}

\subsubsection{Supervised merging process}
 
According to the description in~\ref{sec:principle}, $\phi_i$ are computed on $ \mathcal{G}_{L} $ to discover $ \mathcal{E}_{cal}$ links, and then used to merge rankings on $ \mathcal{G}_{T} $ to discover $\mathcal{E}_{perf} $ links.
Note that we apply the scaling factor $ f \simeq 1.5 $, according to its definition in~\ref{desc-test-phase}, to adapt the $\phi_i$ learnt to the test rankings.
%
%
Considering the value of $g$, it is fixed in order to obtain the best possible performance on the learning set.
Our numerical experiments show that the performance of the algorithm is robust over a large range of values of $g$ (see Table~\ref{tab:rankings}).
It is of course a desirable property, because if the results were varying a lot under small variations of $g$, it would make the parameter $g$ difficult to tune.
Concerning feature selection, we argue in the following that the user may aggregate as many rankings as possible.
In fact, the information provided by different rankings may be redundant, but the merging process is such that the addition of a supplementary ranking is computationally cheap.
Moreover, if a ranking does not bring additional information, it should simply be ignored during the learning process.

\subsubsection{Results}

\begin{figure}[h!]
\begin{center}
\includegraphics[angle=-90,width=0.495\linewidth]{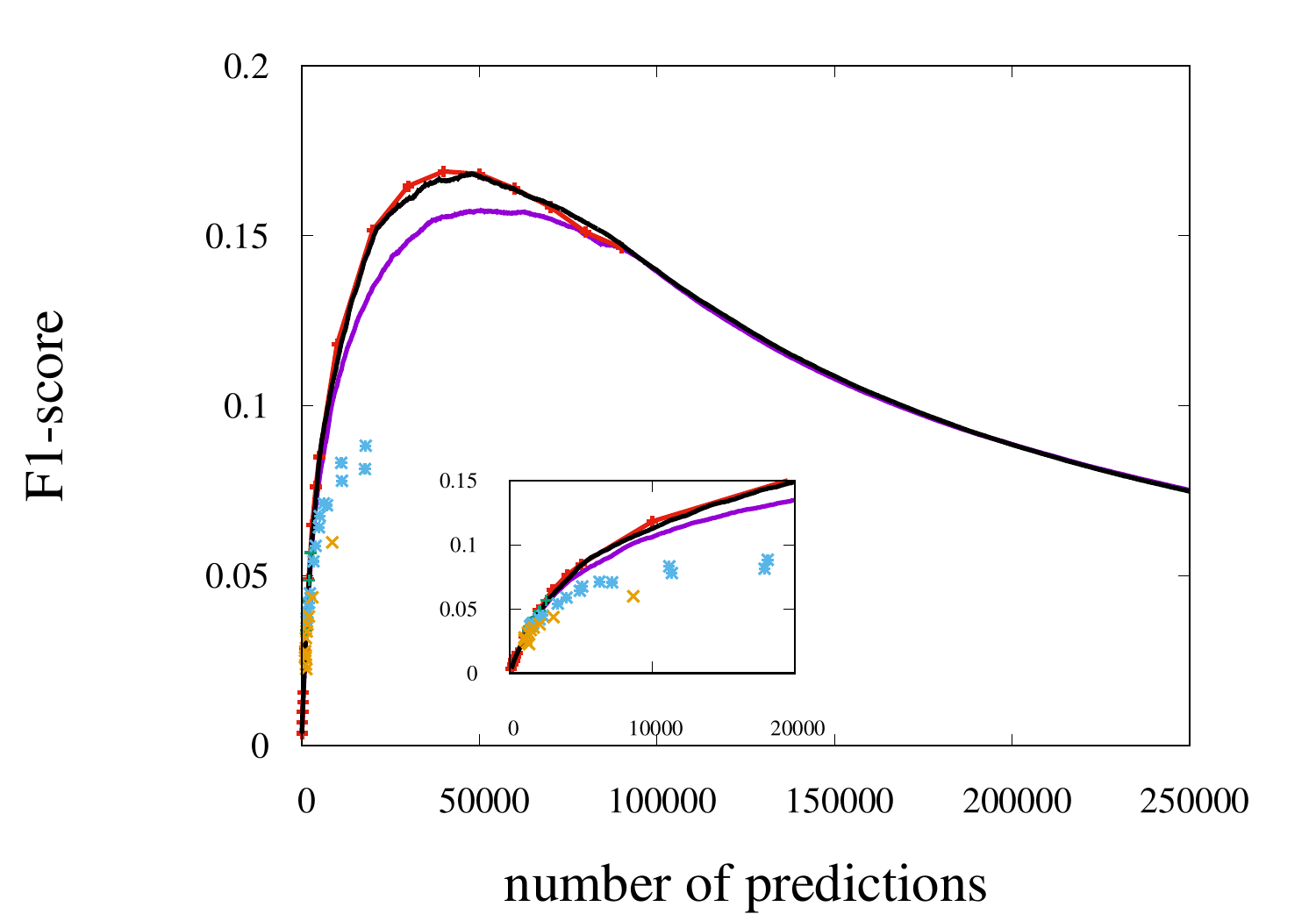}
\includegraphics[angle=-90,width=0.495\linewidth]{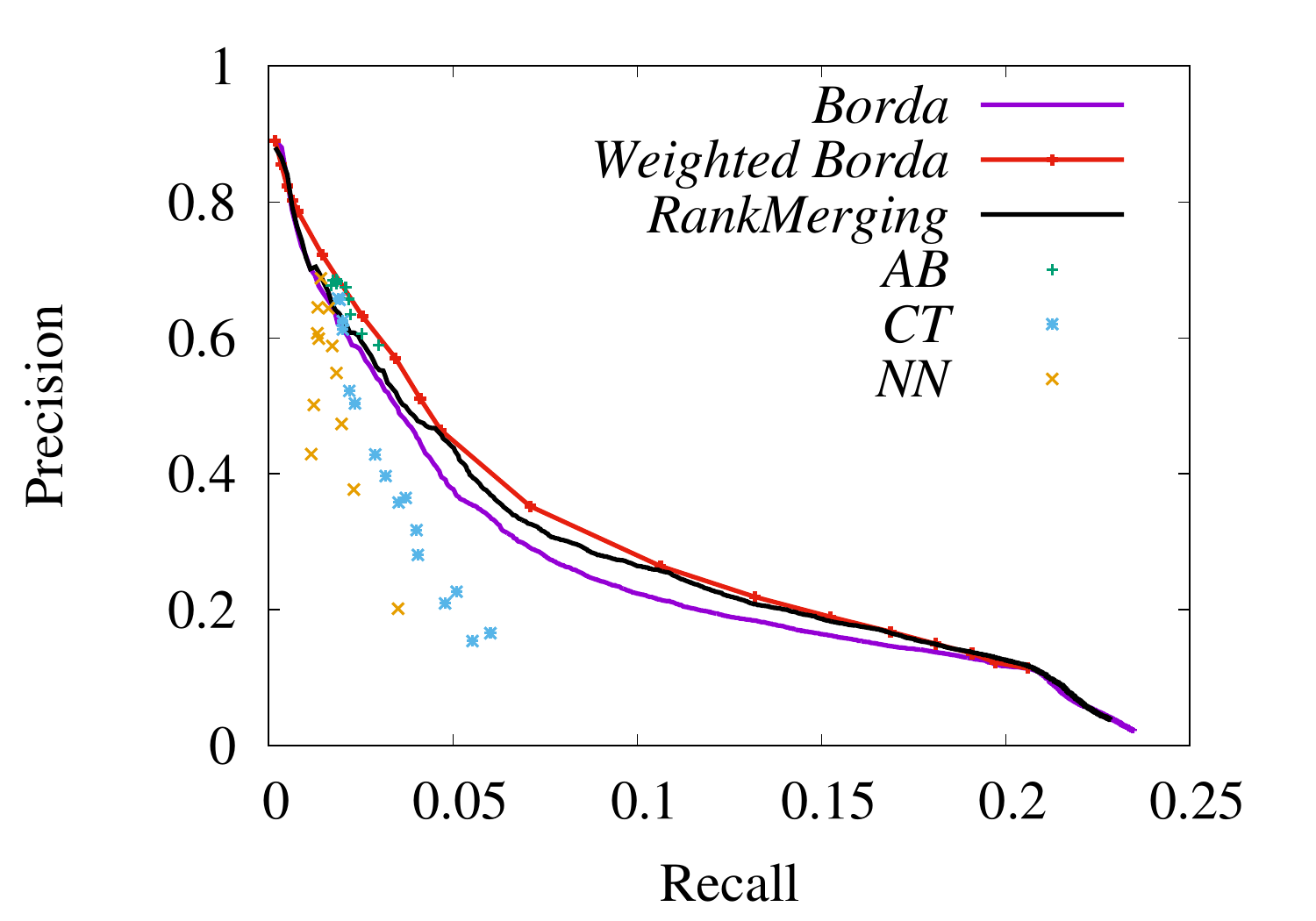}
\end{center}
\caption{\label{fig:res}
Results obtained with \textit{RankMerging} on the test set (learning with $g=200$), compared to various benchmarks.
Left: F1-score as a function of the number of predictions. 
Right: precision versus recall curves.}
\end{figure}

We plot in Figure~\ref{fig:res} the evolution of the F1-score and the precision-recall curve obtained with \textit{RankMerging}, for $g=200$, aggregating the rankings of the following classifiers: \textit{AA}$_w$, \textit{CN}$_w$, $CN$, \textit{SR}$_w$, \textit{Katz}$_w$ ($\gamma=0.1$), \textit{PA}$_w$, \textit{RWR}$_w$ ($p=0.8$) and Borda's method applied to these seven classifiers.
We observe that \textit{RankMerging} performs better than Borda, and consequently better than all the unsupervised methods explored, especially for intermediary recall values.
It was expected, as \textit{RankMerging} incorporates the information of Borda's aggregation here.
%
%
We measure the area under the precision-recall curves to quantify the performances with a scalar quantity.
\textit{RankMerging} increases the area by 8.3\% compared to Borda.
{\modif
Note also that this improvement is not significantly affected by the fact that the method is not deterministic.
Let us recall that there are two sources of non-determinism in the protocol: pairs of the input rankings with similar scores are ranked randomly, and ties between rankers which have similar $ \chi $ during the aggregation process are also broken randomly.
Here, the fluctuations between two realizations of a same experiment are marginal.}

Concerning the supervised classification benchmarks, we observe that they perform well, but only for a low number of predictions (it is comparable to Borda for approximately 1000 to 2000 predictions).
Unsurprisingly, \textit{AdaBoost} is an ensemble method and outperforms \textit{Decision Trees} and \textit{k Nearest Neighbors} for an optimal parameter choice, but the performances are of the same order of magnitude, in line with the observations in~\cite{al2006link}.
As formerly stated, these methods are not designed to cover a wide range of the precision-recall space, and therefore perform very poorly out of their optimal region of use.

On the minus side, \textit{Weighted Borda} method slightly outperforms \textit{RankMerging} on this dataset.
Indeed, we measure that \textit{Weighted Borda} improves by 4.0\% the area under the precision-recall curve compared to \textit{RankMerging}.
We discuss in the next series of experiments, where our method is more accurate, what can be the reasons for this observation.
Another negative point that should be noticed is that \textit{RankMerging} has been designed for problems with large rankings.
The window size $g$ implies an averaging effect which causes the method to lack efficiency on the top-ranked items.
%
As a consequence, it is not suited to problems with low number of predictions -- as would be the case for information retrieval tasks for example.

\paragraph{Feature selection.}

We evaluate the influence of the structural metrics in Table~\ref{tab:rankings}.
A comprehensive study of the matter would be long and repetitive considering all possible combinations of rankings, and we restrict ourselves to a few examples.
An important result to notice here is that the addition of a ranking does not decrease the quality of the merging process -- except for small fluctuations.
So ideally, a user may add any ranking to his own implementation of the method, whatever the source of information is, it should improve -- or at least not worsen -- the prediction.
This statement in consistent with the experimental observations that we have made on all the datasets presented here.
Note that the design of the method is supposed to favour this behaviour: if a ranking brings too little information to the aggregation process, then it should be ignored during the learning process.
The issue of feature selection is critical for the link prediction problem, for instance consensus methods performances drop dramatically with a poor feature choice, so these experimental observations are encouraging for future usages of \textit{RankMerging}.

\begin{table}[h!]
\newcolumntype{M}[1]{>{\centering}m{#1}}
\begin{center}
\begin{tabular}{|M{0.7cm}M{0.7cm}M{0.7cm}M{0.7cm}M{0.7cm}M{0.7cm}M{0.7cm}M{0.7cm}|c|}
\hline
{\tiny \textit{AA}$_w$} & {\tiny \textit{CN}$_w$} & {\tiny \textit{CN}} & {\tiny \textit{SR}$_w$} & {\tiny \textit{Katz}$_w$} & {\tiny \textit{PA}$_w$}  & {\tiny \textit{RWR}$_w$} & {\tiny \textit{Borda}} & {\tiny imp.(\%)}\\
\hline
 x & x & x & x & x & x & x & x & {\scriptsize 8.3} \\
 \hline
 x & x & x & x & x & x & x &  & {\scriptsize 3.2}\\
 x & x & x & x & x & x &   &  & {\scriptsize -0.7}\\
 x & x & x & x & x &   &   &  & {\scriptsize -1.0}\\
 x & x & x & x &   &   &   &  & {\scriptsize -2.0}\\  
\hline
   & x & x & x & x & x & x & x & {\scriptsize 8.2}\\
   &   & x & x & x & x & x & x & {\scriptsize 8.1}\\
   &   &   & x & x & x & x & x & {\scriptsize 3.7}\\  
   &   &   &   & x & x & x & x & {\scriptsize 3.8}\\   
\hline
\end{tabular}
\caption{Improvement (in \%) to Borda's method of the area under the curve in the precision-recall space, for the aggregation of different rankings.
\label{tab:rankings}}
\end{center}
\end{table}

\paragraph{Parameter g.}

The dependency on $g$ value is shown in Table~\ref{tab:g}. 
Results indicate that the performances are close to the maximum in the interval $ [100;300] $ on the learning set.
%
%
Windows of width $g$ are defined during the learning process, as seen in~\ref{sec:principle}.
However, it is interesting to define an equivalent $g_T$ of $g$ on the test set, that is to say to count the number of links present in the next $g_T$ available pairs of the rankings, and see which would be the corresponding improvement.
It allows, indeed, to validate the assumption that the weights computed on the learning set may be extrapolated to the test set.
As confirmed by the results in Table~\ref{tab:g}, the performances are as well close to the maximum in the interval $ [100;300] $ on the test set.
It confirms that choosing $g$ according to the best aggregation on the learning set ensures that the final performance of \textit{RankMerging} should be good.
It is interesting to note that the performance is maximum for these intermediate values, around $g=200$.
This is the right balance between small $g$ which fluctuate too much to provide a good local optimum, and large $g$, for which the windows are too large to properly evaluate which ranking is the best ranker available, as discussed in Section~\ref{sec:opt_justif}.

\begin{table}[h!]

\newcolumntype{M}[1]{>{\centering}m{#1}}

\begin{small}
\begin{center}
{
\begin{tabular}{|c|M{7mm}|c|c|c|c|c|c|c|} 
\hline
$g$ & 10 & 100 & 200 & 300 & 400 & 500 & 1000 & 2000  \\
\hline
imp. & -0.8 &  5.5 & 5.4 & 5.2 & 5.0 & 4.7 & 4.0 & 2.7 \\
\hline
\end{tabular}

\vspace{2mm}

\begin{tabular}{|c|M{7mm}|c|c|c|c|c|c|c|} 
\hline
$g_T$ & 10 & 100 & 200 & 300 & 400 & 500 & 1000 & 2000  \\
\hline
imp. & 2.7 & 8.2 & 8.3 & 7.9 & 7.4 & 7.2 & 6.4 & 5.6 \\
\hline
\end{tabular}
}
\end{center}
\end{small}
\caption{Improvement to Borda's method of the area under the curve in the precision-recall space, for different values of $g$.
Top: learning set, bottom: test set.
\label{tab:g}}
\end{table}

\subsection{Series 2}

In the second series of experiments, we focus on DBLP and Pokec datasets.
Here, we investigate in more details the impact of the sizes of the learning and test sets.

\subsubsection{Protocol}

A few points differ from the protocol of the first series, in order to easily change the sizes of the learning and test sets. 
All the nodes belong to both $\mathcal{G}_{L}$ and $\mathcal{G}_{T}$, the partition is made on the set of links $ \mathcal{E} $.
We use the same denomination, that is to say:
\begin{itemize}
\item $\mathcal{E}_{L}$ are the links of $\mathcal{G}_{L}$,
\item $\mathcal{E}_{cal}$ are the links to guess during the learning phase, to calibrate our model,
\item $\mathcal{E}_{perf}$ are the links to guess during the testing phase, to evaluate the performance of the method.
\end{itemize}
During the learning phase, $\mathcal{E}_{L}$ links are used to guess $\mathcal{E}_{cal}$ links.
During the test phase, the links of $\mathcal{G}_{T}$, that is to say $\mathcal{E}_{T} = \mathcal{E}_{L} \cup \mathcal{E}_{cal}$ are used to guess $\mathcal{E}_{perf}$.
We generate several samples such that $|\mathcal{E}_{cal}| = |\mathcal{E}_{perf}|$, but with various values for $ |\mathcal{E}_{cal}| $, the missing information increases as this set grows larger.
The samples are defined by the missing links ratio $  |\mathcal{E}_{cal}| / |\mathcal{E}| $.

Another noteworthy difference is that these networks have a much higher average degree than the PSP network, making the computation of the large-scale rankers expensive in both memory and time.
Therefore, we limited ourselves to the less costly local and intermediary metrics, more precisely: \textit{AA}$_w$, \textit{CN}$_w$, \textit{SR}$_w$, \textit{RA}$_w$, \textit{LP}$_w$ ($\gamma = 0.1$) and Borda's aggregation.

For the same reasons, the number of pairs is very large when only considering nodes at distance~2 (larger than $ 10^8$).
{\modif
Since it would not make much sense to predict that many links in a social network, we choose to reduce the rankings length by focusing on the intersection of rankings: we kept only $10^6$ pairs, which feature the intersection of the tops of all rankings.
Borda's method is then applied on the intersected rankings.}
In this setting of experiments, the rankings in the learning set and in the test set have the same sizes so that the scaling factor $f=1$ during the whole series.

\subsubsection{Results}

\begin{figure}[h!]
\begin{center}
\includegraphics[angle=-90,width=0.49\linewidth]{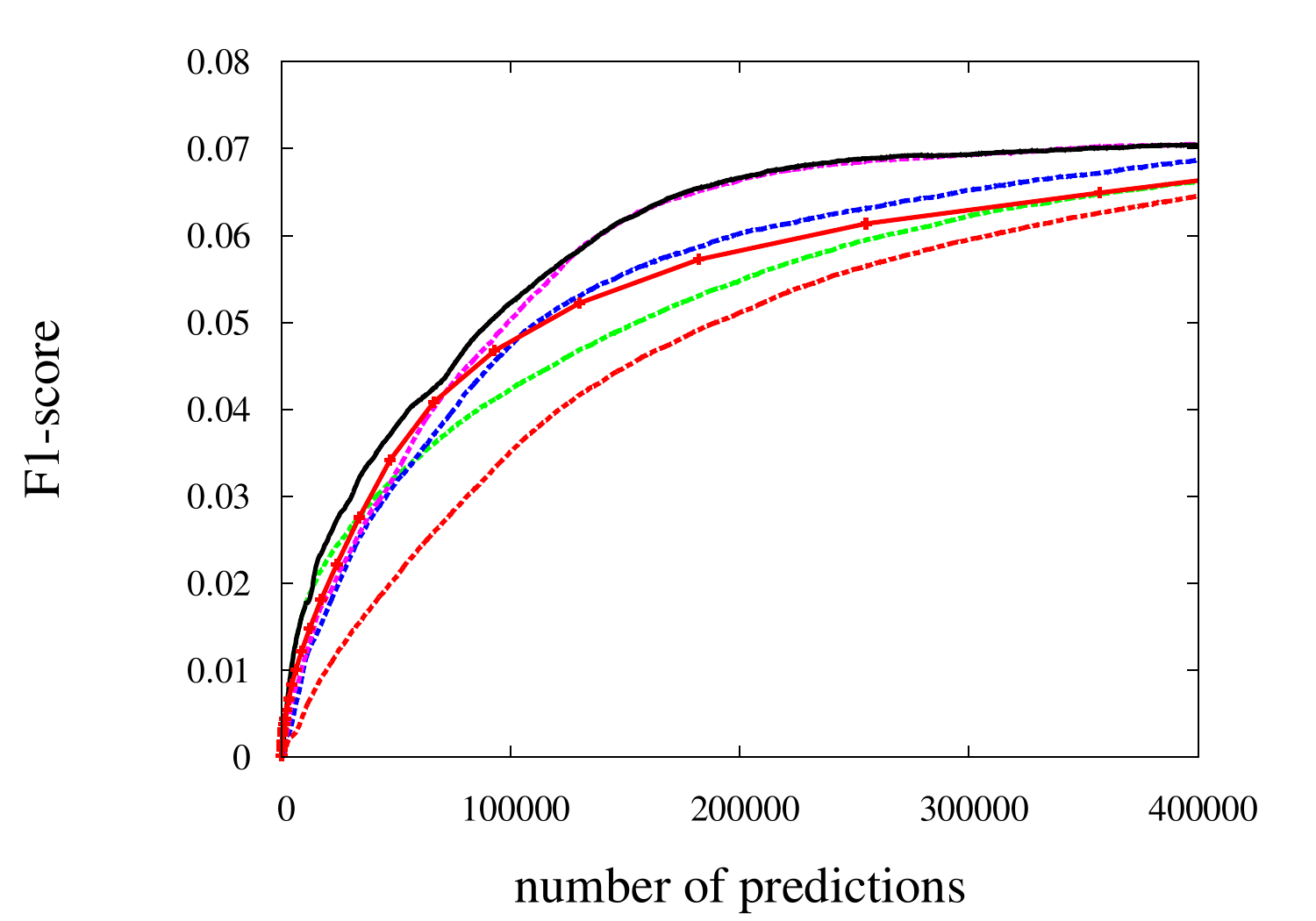}
\includegraphics[angle=-90,width=0.49\linewidth]{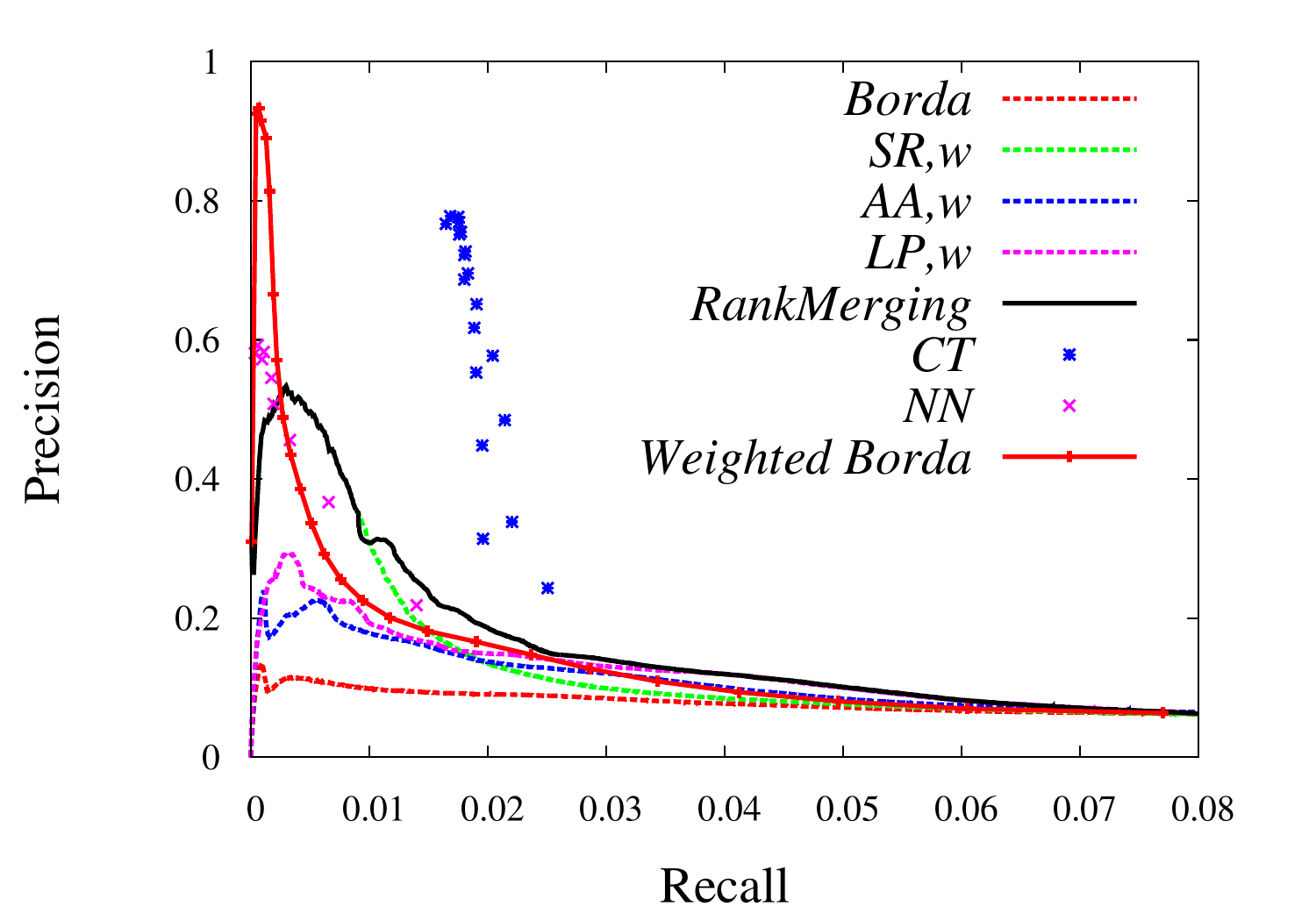}
\includegraphics[angle=-90,width=0.49\linewidth]{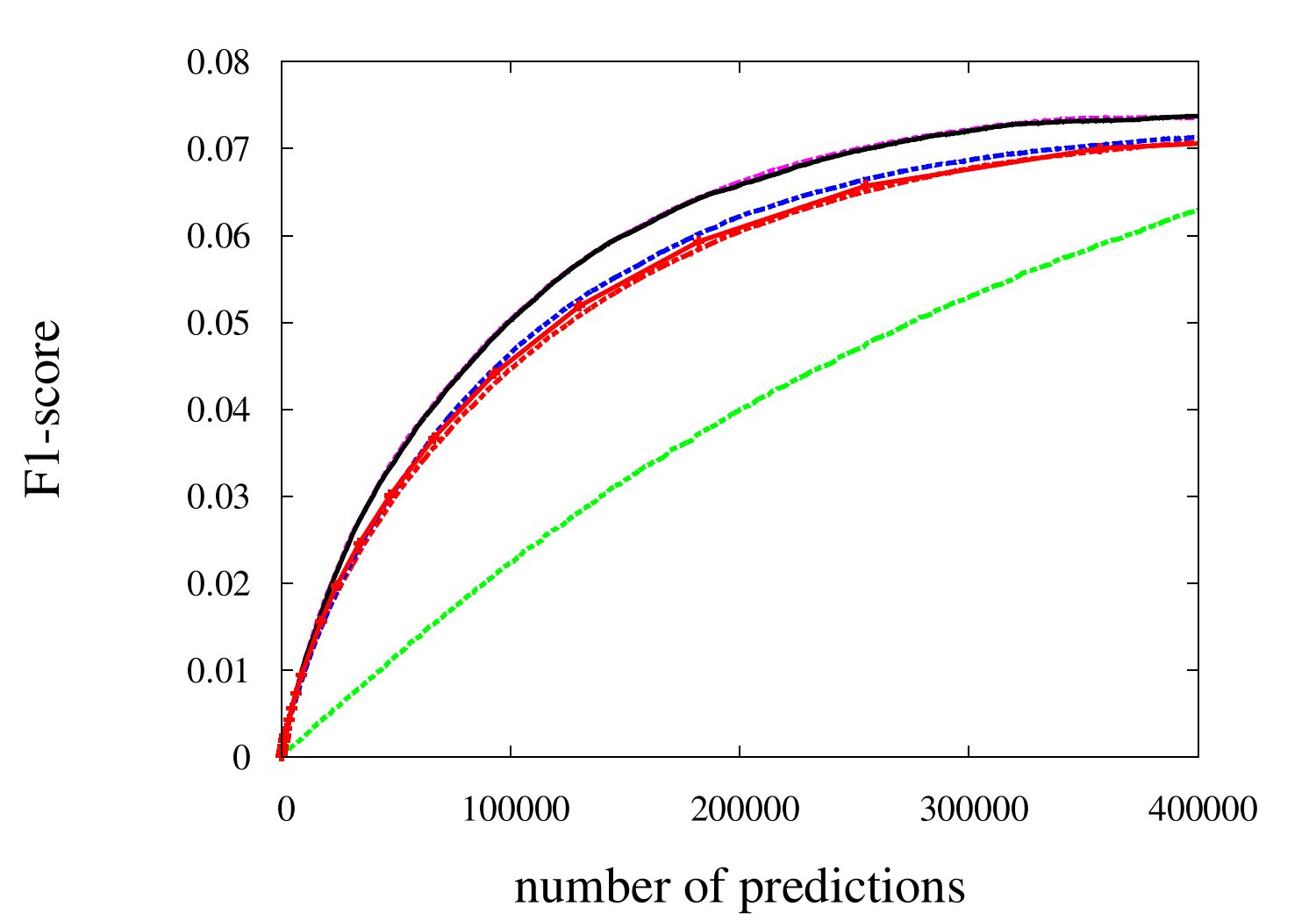}
\includegraphics[angle=-90,width=0.49\linewidth]{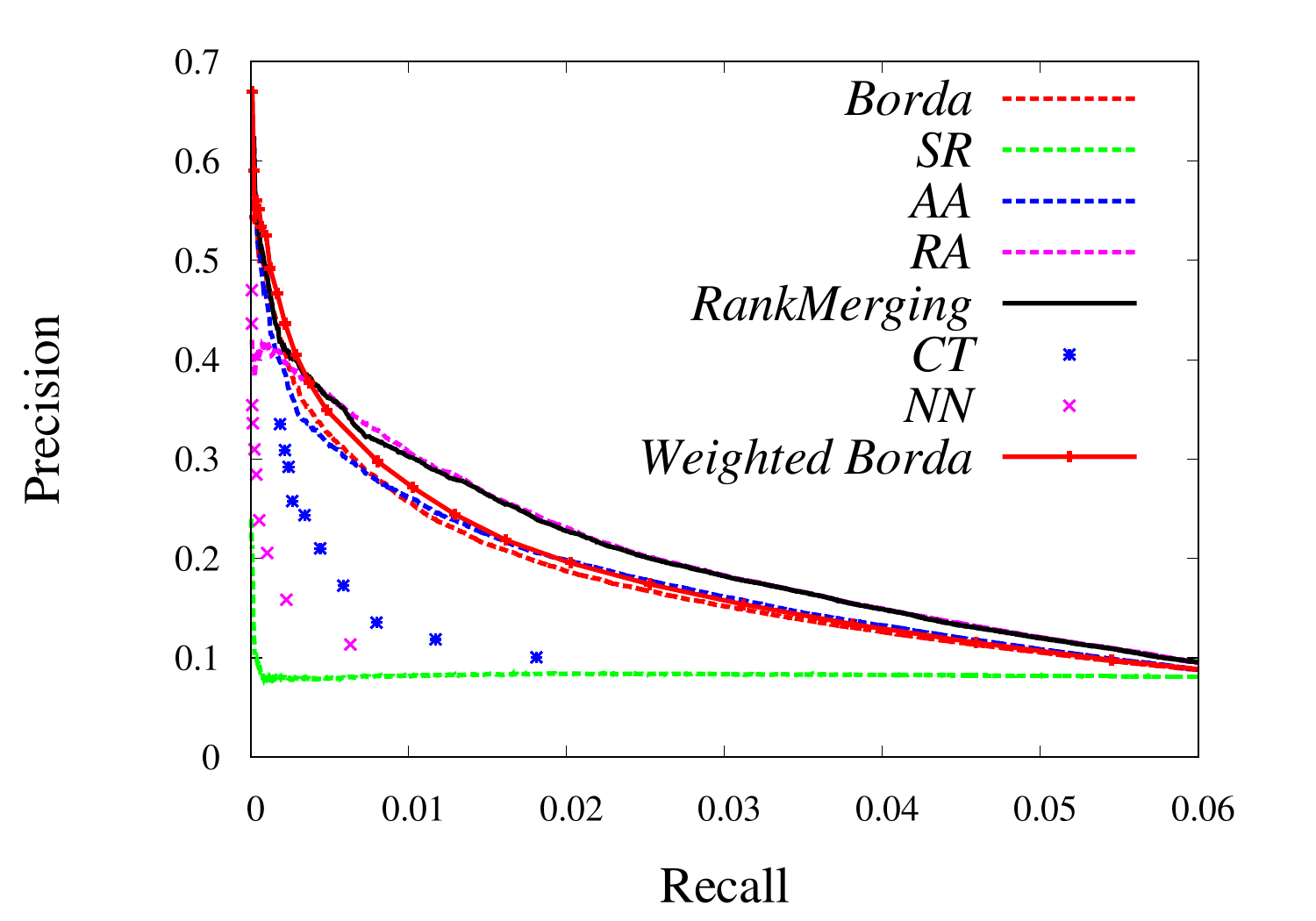}
\end{center}
\caption{\label{fig:res_pokec-dblp}
Results obtained with \textit{RankMerging} on the test set.
Top: DBLP (learning with $g=4000$).
Bottom: Pokec (learning with $g=2000$). 
}
\end{figure}

We plot in Figure~\ref{fig:res_pokec-dblp} the results obtained on the test sets of both DBLP and Pokec datasets, for samples where the missing links ratio $ |\mathcal{E}_{cal}| / |\mathcal{E}| = 0.077$.
In both cases, \textit{RankMerging} outperforms the unsupervised methods, but the improvement is much more visible for DBLP than for Pokec.
In this series of experiments too, performances are not significantly affected by the fact that the method is not deterministic.

In the case of DBLP, a closer examination shows that at the beginning of the process, we closely follow S\o rensen index curve, which is the most efficient unsupervised ranking for low number of predictions.
Then its performance drops, and other rankings are chosen during the aggregation process (mostly Local Path ranking).
In the case of Pokec, pairs aggregated initially mainly come from Adamic-Adar ranking, then this index soon gets less efficient than Resource Allocation index, which takes over until the end of the merging process.
Pokec gives a good indication of what happens when the rankings are not complementary enough, and that one of them is more efficient than the others: the aggregation nearly always choose the pairs from this ranking and do not improve significantly the performance.
Notice that in both cases, Borda's method is not the most efficient unsupervised method: as some rankings perform very poorly, they dramatically hinder the performance of Borda's method.

The comparison to supervised classification methods is interesting here\footnote{Notice that \textit{AdaBoost} does not provide exploitable results on both networks as it predicts too few links.}: \textit{Classification Trees} as well as \textit{k Nearest Neighbors} perform poorly on Pokec, however the results on DBLP show that \textit{NN} and above all \textit{CT} perform better than \textit{RankMerging} on a limited range. 
This observation highlights a limit of our learning-to-rank aggregation method for the purpose of link prediction: the prediction performance cannot be widely better than the performances of the rankings that have been used as features for the learning, while classification methods can.
There is however a compensation, as the number of predictions is not constrained, as in the case of classification.

\begin{figure}[h!]
\begin{center}
\includegraphics[width=0.65\linewidth]{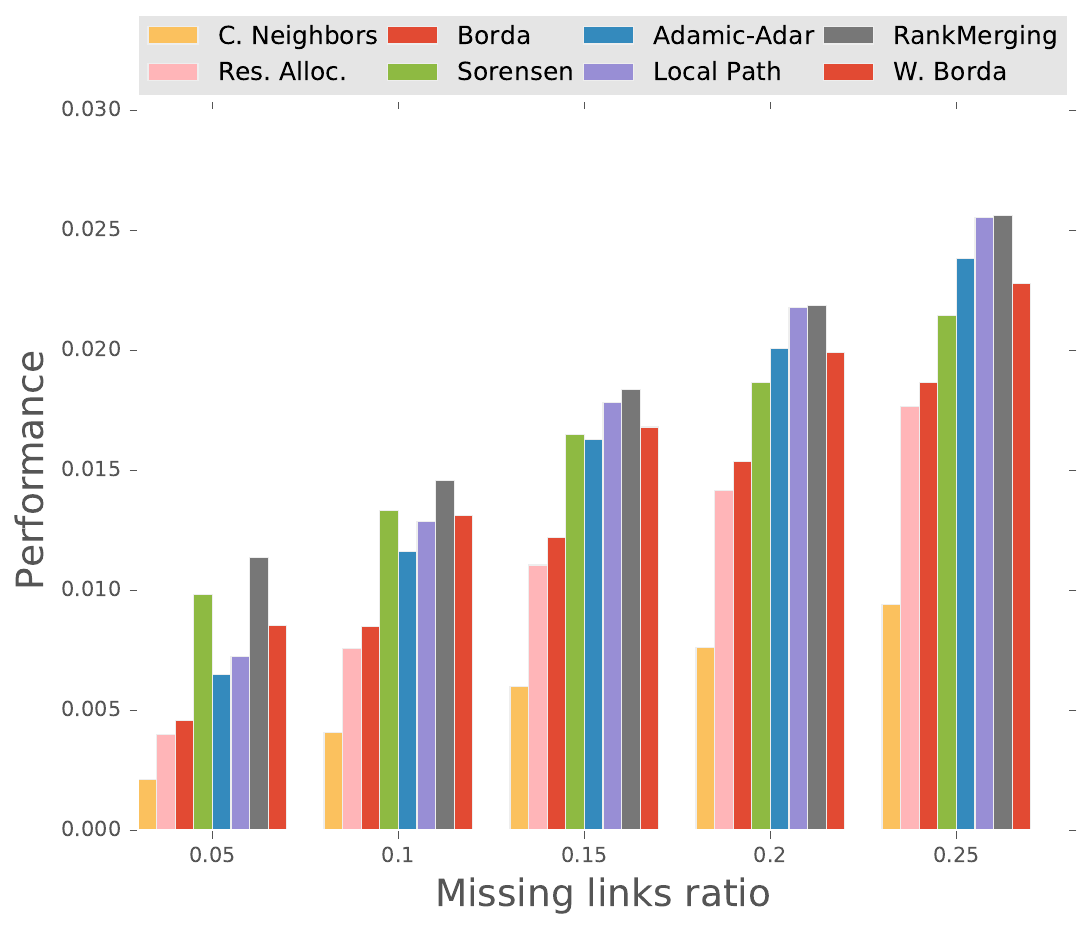}
\includegraphics[width=0.65\linewidth]{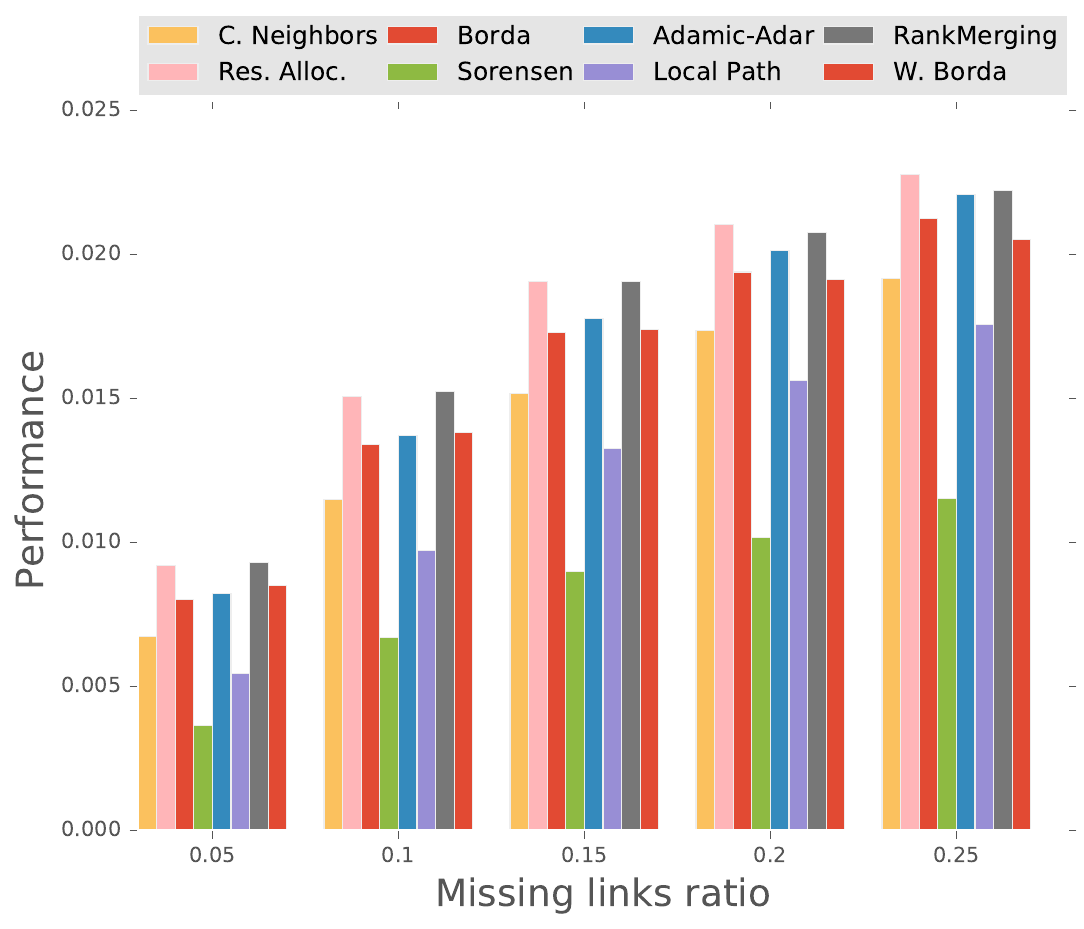}
\end{center}
\caption{\label{fig:samples_pokec-dblp}
Area under the precision-recall curve for different missing links ratios.
Top: DBLP dataset, bottom: Pokec dataset.}
\end{figure}

We also compare to the \textit{Weighted Borda} method, which is the only learning-to-rank supervised method considered here, given the very large sizes of the rankings (c.f.~\ref{subsec:bench}).
To implement this method, we need to learn a new set of coefficients for every ranking size, that is to say for every measurement point in the precision-recall space.
We observe that the \textit{Weighted Borda method} is efficient for a relatively low number of predictions.
It outperforms \textit{RankMerging} on a short range: on $ [0;2000] $ on DBLP and on $ [0;5000] $ on Pokec.
But as in the case of supervised classification methods, its performance drops for larger number of predictions and \textit{RankMerging} is more efficient when considering the area under the precision-recall curve, as can be seen in Figure~\ref{fig:samples_pokec-dblp}.
This result contrasts with the observation in the first series of experiments where \textit{Weighted Borda} is more efficient.
We think that it stems from the fact that in the second series of experiments, the unsupervised rankings are more redundant, in the sense that different rankers rank the same items high.
While \textit{RankMerging} takes into account this property to compute the contribution of a ranking during the learning phase, \textit{Weighted Borda} does not.
%

\paragraph{Learning set size.}

We now explore the impact of the learning set size on the performance of the method.
We generated five different samples with missing links ratios: 0.05, 0.10, 0.15, 0.20 and 0.25.
In Figure~\ref{fig:samples_pokec-dblp}, we plot the area under the precision-recall curves for the samples generated, and compare them to the most efficient unsupervised methods tested.
We observe that \textit{RankMerging} outperforms both the unsupervised and supervised learning-to-rank methods in nearly all cases.
However, the differences vanish when the size of the learning set decrease, that is to say when the missing information grows.
It seems that this observation stems from the fact that a ranker dominates the others when information is missing, so that the merging method tends to stick to the most performing ranker.

{\modif
\subsection{Series 3}

In the third series of experiments, we focus on the Facebook dataset, which is considered in a predictive context.
%

\subsubsection{Protocol}

In the case of this series, we aim at predicting links, which implies that the protocol is adapted to such a task.
\begin{itemize}
\item The graph $ \mathcal{G} _L $ contains the 46,953 nodes which appear at least once during the whole period of collection (from January $1^{st}$ 2007 to November $ 30^{th} $ 2008), that is the set $\mathcal{V}$.
Its links $ \mathcal{E} _L $ are the 304,306 links which appear in the dataset during the first 600 days period.
\item Links used for calibration $ \mathcal{E} _{cal} $ are the links appearing for the first time between day 601 and day 650, that is 16,411 links.
\item And the links to be guessed during the test phase $ \mathcal{E} _{perf} $ are the links appearing for the first time between day 651 and day 700, that is 22,509 links.
\end{itemize}

As usual, $ \mathcal{G} _L = (\mathcal{V},\mathcal{E} _L) $ is used to label pairs during the learning phase in order to guess links in $ \mathcal{E} _{cal} $.
Then the graph $ \mathcal{G} _T = (\mathcal{V},\mathcal{E} _L \cup \mathcal{E} _{cal}) $ is used to label pairs during the test phase to guess links in $ \mathcal{E} _{perf} $.

The ranking metrics used to produce the unsupervised input rankings are $ CN_{w}$, $AA_{w}$, $RA_{w}$, $SR_{w}$, $LP$ (with $ \gamma = 0.1 $) and the unsupervised \textit{Borda} aggregation of these rankings.
We compare the results obtained to the \textit{supervised Borda} method (with weights computed from the precision of the input rankings).
Note also that we have dropped classification techniques for comparison purposes to the other methods, as it has been discussed previously that the prediction yielded have very different properties compared to the other techniques.

\subsubsection{Results}

The results obtained are gathered in Figure~\ref{fig:fb-results}.
We can see that the overall aspect of the results are qualitatively similar to what we have observed in the other series of experiments. 
When computing the area under the precision-recall curve, \textit{RankMerging} outperforms by 6.6\% the unsupervised version of \textit{Borda} aggregation, which is the second most efficient method here, and  by 8.1\% \textit{supervised Borda}, which is the third most efficient.
A closer look at the learning process indicates that \textit{RankMerging} first selects pairs essentially from the \textit{Borda} ranking, then jumps from \textit{Borda} to the \textit{Adamic-Adar} ranking regularly.
It starts selecting pairs from other rankings (\textit{Resource Allocation}) only after the first 100,000 predictions.
These experiments thus show that the method is also efficient in a prediction context when compared to other methods.
However, it should be noticed that precisions in a prediction context are low for all methods which comes from the fact that this task is even more challenging than link recovery.
\begin{figure}[!h]
\begin{center}
\includegraphics[angle=-90,width=0.495\linewidth]{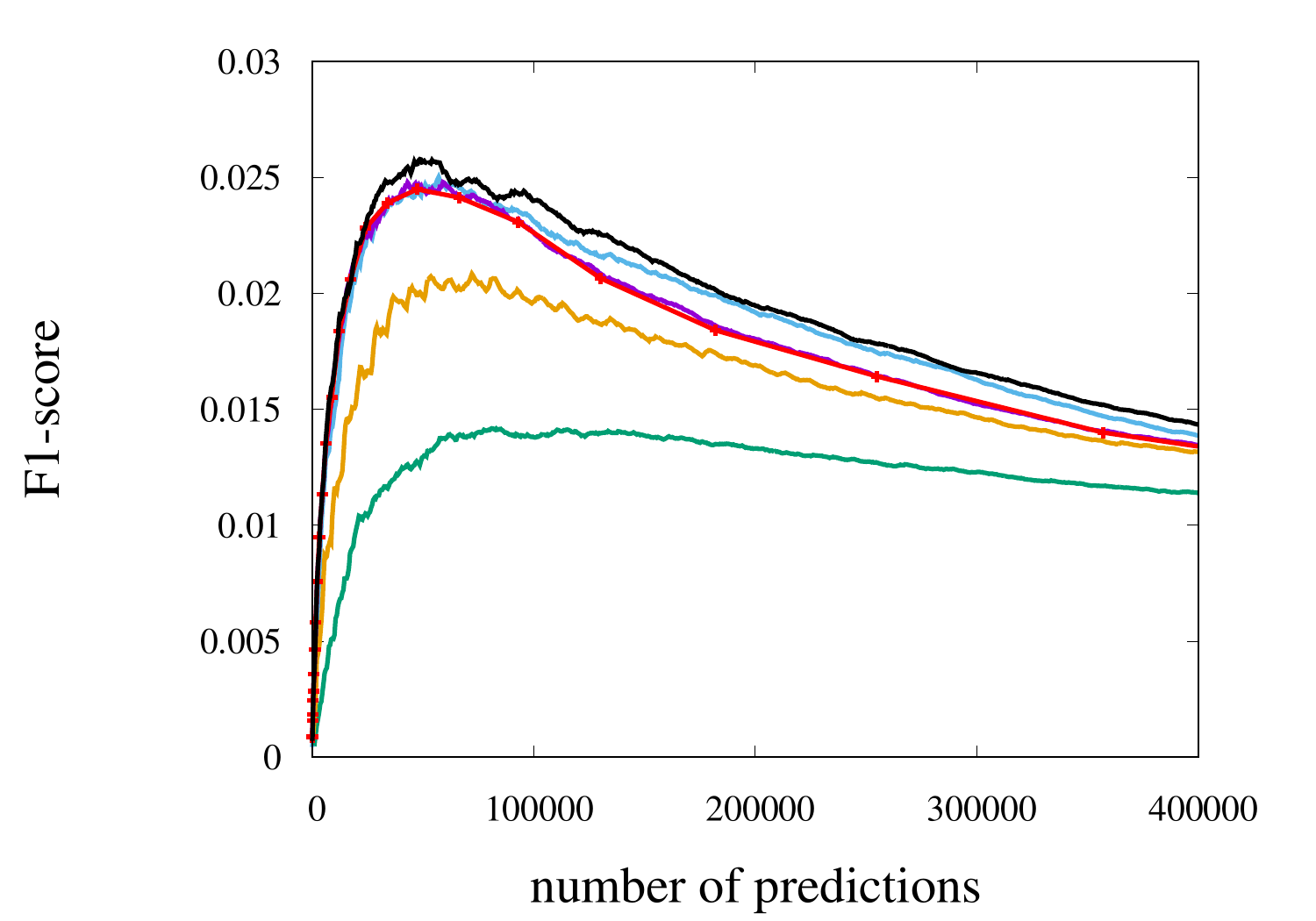}
\includegraphics[angle=-90,width=0.495\linewidth]{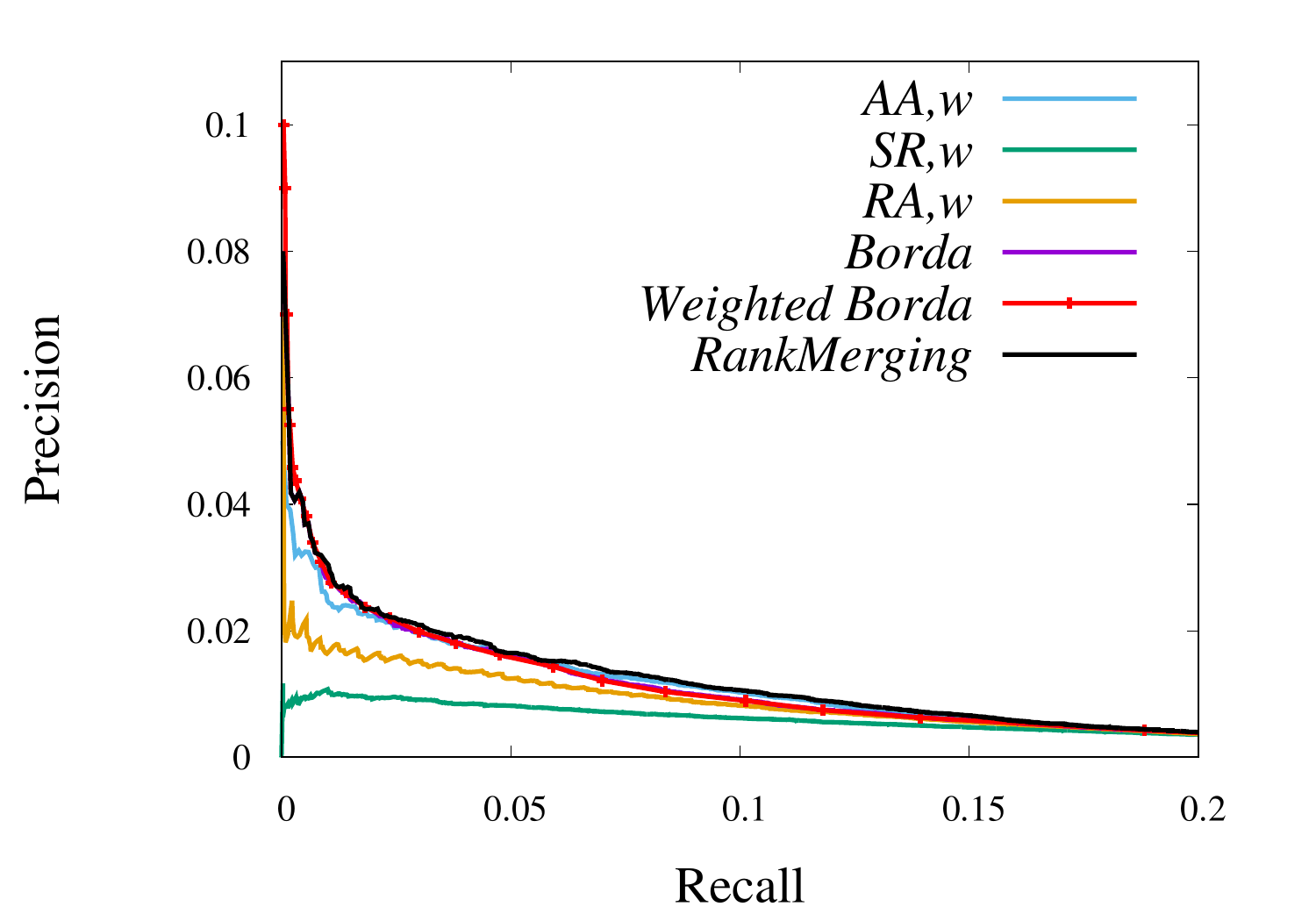}
\end{center}
\caption{\label{fig:fb-results}
Results obtained on the link prediction problem on Facebook data.
Left: F1-score as a function of the number of predictions. 
Right: precision versus recall curves.}
\end{figure}


%
}

\subsection{Experimental running times.}\label{exp-running-times}
To give the reader a better grasp of the practical running times, we indicate in Table~\ref{tab-running-times} the order of magnitude of the running times of the two first series of experiments.
We used standard implementations on a workstation with 16$\cdot$3~GHz CPU throughout the experiments and specify the order of magnitude of the computation times. 
In all cases, we report the total time to obtain the results presented in the series of experiments, meaning that we sum the running times of the experiments of the learning and test phases for the supervised processes\footnote{We do not include in the computation times of the supervised methods the time necessary to produce the unsupervised rankings inputs.}. 
Running times are measured in seconds and given with one or two significant digits.
The results reported here correspond to:
\begin{itemize}
\item Unsupervised rankings: we report the execution times of a typical local feature (\textit{Adamic-Adar})\footnote{The running times are of the same magnitude for the other local metrics.}, an intermediary feature (\textit{Local Path}) and \textit{Borda} method.
\item Supervised classification: we report the execution times for the three classification methods used (\textit{Classification Trees, k Nearest Neighbours, AdaBoost}).
As each measurement point needs a calibration of the model, the results reported correspond to a 10 points measurement.
\item Supervised learning-to-rank: we report the execution times of the \textit{Weighted Borda} method and \textit{RankMerging}. 
Note that the computation time of the \textit{Weighted Borda} method depends on the number of measurement points as a new point demands a new aggregation with different weights.
Similarly, one has to adjust the parameter $g$ of \textit{RankMerging} using its best perfomance on the learning set, therefore we need to run the learning phase of the method with several $g$ values.
In both cases, we use 24 measurements to evaluate the running times.
%
%
\end{itemize}
%


\begin{table}[!h]
\begin{center}
\begin{footnotesize}
\begin{tabular}{|c|c|c|c||c|c|c||c|c|}
\hline
& \multicolumn{3}{c||}{Unsupervised} & \multicolumn{3}{c||}{Sup. Classification} & \textit{Weighted} &  \textit{Rank} \\
& Local & Inter & Borda & CT & NN & AB &  \textit{Borda}  & \textit{Merging}\\
\hline
PSP   & $< 10$           & $30$             & $< 10$ & $ 210 $ & $4\cdot10^4$ & $ 500 $  & $ 200 $  &  $200$ \\
DBLP  & $ 6 \cdot 10^4 $ & $ 7 \cdot 10^4 $ & $ 40 $ & $ 400 $ & $ 800 $      & $ 1100 $ & $ 1200 $ &  $1200$ \\
Pokec & $ 5 \cdot 10^4 $ & $ 1 \cdot 10^5 $ & $ 30 $ & $ 300 $ & $ 400 $      & $ 1000 $ & $ 1000 $ &  $1000$ \\
%
%
\hline
\end{tabular}
\end{footnotesize}
\end{center}
\caption{\label{tab-running-times}
Running times of the different prediction models, given in seconds. 
}
\end{table}

Several points may be noticed regarding these measures.
First, \textit{RankMerging} and \textit{Weighted Borda} have comparable running times in all cases, which was expected as \textit{RankMerging}, like \textit{Weighted Borda}, go through each ranking exactly once.
They are both suited to supervised learning-to-rank on large rankings.

Considering the whole learning-to-rank process, the bottleneck is the computation of the unsupervised rankings.
Concerning PSP, it can be observed that the running times of local unsupervised methods is only a few dozens of seconds while it is several hours on the Pokec and DBLP datasets. 
It stems from the fact that PSP is much sparser than DBLP and Pokec, and the typical sizes of the rankings are shorter for PSP: as mentioned previously, PSP rankings are of the order of $ 10^5 $ items, while DBLP and Pokec rankings are larger than $ 10^8 $ items.
Note also that the computation times of global metrics (not reported here) is much larger than local metrics as they generally rely on the computation of matrix products.
For example, \textit{RWR} typically takes a few hours to run on PSP, while it is untractable on Pokec and DBLP.

\section{Conclusion}


In this work, we have presented \textit{RankMerging}, a supervised method which aggregate rankings from unsupervised rankers to improve the performance of link prediction.
This learning-to-rank method is straightforward and computationally cheap -- its complexity is $ O(\alpha \cdot \theta)$, where $\alpha$ is the number of rankings aggregated and $\theta$ the number of predictions.
It is suited to prediction in large networks, as $\theta$ can be tuned according to the application needs.
Its design implies that the precision on top-ranked items cannot be significantly improved, making \textit{RankMerging} more appropriate for relatively large number of predictions.
%
%
%
We implemented it on three different large social networks, and showed that in each case, it competes with the methods available at this scale, such as the \textit{Weighted Borda} method.
It is especially efficient when the unsupervised input rankers are complementary, but its performances are robust to the addition of redundant information.

So far, we have exclusively focused on structural information in order to predict unknown links.
However, the framework is general and any feature providing a ranking for likely pairs of nodes can be incorporated.
Additional structural classifiers are an option, but other types of attributes can also be considered, such as the profile of the users (age, hometown, etc), or timings of their interactions.
In the latter case, for instance, if $i$ and $j$ are both interacting with $k$ within a short span of time, it is probably an indication of a connection between $i$ and $j$.
From a theoretical perspective, \textit{RankMerging} provides a way to uncover the mechanisms of link creation, by identifying  which sources of information play a dominant role in the quality of a prediction.
The method could be applied to various types of networks, especially when links are difficult to detect.
Applications include network security -- for example by detecting the existence of connections between machines of a botnet -- and biomedical engineering -- for screening combinations of active compounds and experimental environments in the purpose of medicine discovery.

\section*{Acknowledgements}

The authors would like to thank Emmanuel Viennet and Maximilien Danisch for useful bibliographic indications.
L.T. also thanks Dan Timsit for his feedback and implementation of the C++ version of the algorithm.

This paper presents research results of the Belgian Network DYSCO (Dynamical Systems, Control, and Optimization), funded by the Interuniversity Attraction Poles Programme, initiated by the Belgian State, Science Policy Office. The scientific responsibility rests with its authors.
%
%
This work is also funded in part by the ANR (French National Agency of Research) under grants ANR-15-CE38-0001 (AlgoDiv) and ANR-13-CORD-0017-01 (CODDDE), by the French program "PIA - Usages, services et contenus innovants" under grant O18062-44430 (REQUEST), and by the Ile-de-France FUI21 program under grant 16010629 (iTRAC).
We also acknowledge support from FNRS.


\bibliographystyle{plainnat}
\bibliography{biblio}

\begin{thebibliography}{41}
\providecommand{\natexlab}[1]{#1}
\providecommand{\url}[1]{\texttt{#1}}
\expandafter\ifx\csname urlstyle\endcsname\relax
  \providecommand{\doi}[1]{doi: #1}\else
  \providecommand{\doi}{doi: \begingroup \urlstyle{rm}\Url}\fi

\bibitem[Al~Hasan and Zaki(2011)]{al2011survey}
M.~Al~Hasan and M.~Zaki.
\newblock A survey of link prediction in social networks.
\newblock In \emph{Social network data analytics}, pages 243--275. Springer,
  2011.

\bibitem[Al~Hasan et~al.(2006)Al~Hasan, Chaoji, Salem, and Zaki]{al2006link}
M.~Al~Hasan, V.~Chaoji, S.~Salem, and M.~Zaki.
\newblock Link prediction using supervised learning.
\newblock In \emph{SDM’06: Workshop on Link Analysis, Counter-terrorism and
  Security}, 2006.

\bibitem[Arrow(2012)]{arrow2012social}
K.J. Arrow.
\newblock \emph{Social choice and individual values}, volume~12.
\newblock Yale university press, 2012.

\bibitem[Backstrom and Leskovec(2011)]{backstrom2011supervised}
L.~Backstrom and J.~Leskovec.
\newblock Supervised random walks: predicting and recommending links in social
  networks.
\newblock In \emph{Proceedings of the fourth ACM international conference on
  Web search and data mining}, pages 635--644. ACM, 2011.

\bibitem[Benchettara et~al.(2010)Benchettara, Kanawati, and
  Rouveirol]{benchettara2010supervised}
N.~Benchettara, R.~Kanawati, and C.~Rouveirol.
\newblock Supervised machine learning applied to link prediction in bipartite
  social networks.
\newblock In \emph{International Conference on Advances in Social Networks
  Analysis and Mining (ASONAM)}, pages 326--330. IEEE, 2010.

\bibitem[Bliss et~al.(2013)Bliss, Frank, Danforth, and
  Dodds]{bliss2013evolutionary}
C.A. Bliss, M.R. Frank, C.M. Danforth, and P.S. Dodds.
\newblock An evolutionary algorithm approach to link prediction in dynamic
  social networks.
\newblock \emph{arXiv:1304.6257}, 2013.

\bibitem[Burges et~al.(2011)Burges, Svore, Bennett, Pastusiak, and
  Wu]{burges2011learning}
C.J.C. Burges, K.M. Svore, P.N. Bennett, A.~Pastusiak, and Q.~Wu.
\newblock Learning to rank using an ensemble of lambda-gradient models.
\newblock \emph{Journal of Machine Learning Research-Proceedings Track},
  14:\penalty0 25--35, 2011.

\bibitem[Cao et~al.(2007)Cao, Qin, Liu, Tsai, and Li]{cao2007learning}
Z.~Cao, T.~Qin, T.Y. Liu, M.F. Tsai, and H.~Li.
\newblock Learning to rank: from pairwise approach to listwise approach.
\newblock In \emph{Proceedings of the 24th international conference on Machine
  learning}, pages 129--136. ACM, 2007.

\bibitem[Chapelle and Keerthi(2010)]{chapelle2010efficient}
O.~Chapelle and S.S. Keerthi.
\newblock Efficient algorithms for ranking with svms.
\newblock \emph{Information Retrieval}, 13\penalty0 (3):\penalty0 201--215,
  2010.

\bibitem[Chapelle et~al.(2011)Chapelle, Chang, and Liu]{chapelle2011future}
O.~Chapelle, Y.~Chang, and T.Y. Liu.
\newblock Future directions in learning to rank.
\newblock In \emph{Yahoo! Learning to Rank Challenge}, pages 91--100, 2011.

\bibitem[Comar et~al.(2011)Comar, Tan, and Jain]{comar2011linkboost}
P.M. Comar, P.N. Tan, and A.K. Jain.
\newblock Linkboost: A novel cost-sensitive boosting framework for
  community-level network link prediction.
\newblock In \emph{11th International Conference on Data Mining (ICDM)}, pages
  131--140. IEEE, 2011.

\bibitem[Dasgupta et~al.(2008)Dasgupta, Singh, Viswanathan, Chakraborty,
  Mukherjea, Nanavati, and Joshi]{dasgupta2008social}
K.~Dasgupta, R.~Singh, B.~Viswanathan, D.~Chakraborty, S.~Mukherjea, A.A.
  Nanavati, and A.~Joshi.
\newblock Social ties and their relevance to churn in mobile telecom networks.
\newblock In \emph{Proceedings of the 11th International Conference on
  Extending Database Technology}, pages 668--677. ACM, 2008.

\bibitem[Davis et~al.(2013)Davis, Lichtenwalter, and
  Chawla]{davis2013supervised}
D.~Davis, R.~Lichtenwalter, and N.V. Chawla.
\newblock Supervised methods for multi-relational link prediction.
\newblock \emph{Social Network Analysis and Mining}, 3\penalty0 (2):\penalty0
  127--141, 2013.

\bibitem[de~Borda(1781)]{deborda1781memoire}
J.C. de~Borda.
\newblock M{\'e}moire sur les {\'e}lections au scrutin.
\newblock 1781.

\bibitem[Dwork et~al.(2001)Dwork, Kumar, Naor, and Sivakumar]{dwork2001rank}
C.~Dwork, R.~Kumar, M.~Naor, and D.~Sivakumar.
\newblock Rank aggregation methods for the web.
\newblock In \emph{Proceedings of the 10th international conference on World
  Wide Web}, pages 613--622. ACM, 2001.

\bibitem[Freund et~al.(2003)Freund, Iyer, Schapire, and
  Singer]{freund2003efficient}
Y.~Freund, R.~Iyer, R.E. Schapire, and Y.~Singer.
\newblock An efficient boosting algorithm for combining preferences.
\newblock \emph{The Journal of machine learning research}, 4:\penalty0
  933--969, 2003.

\bibitem[Herbrich et~al.(1999)Herbrich, Graepel, and
  Obermayer]{herbrich1999large}
R.~Herbrich, T.~Graepel, and K.~Obermayer.
\newblock Large margin rank boundaries for ordinal regression.
\newblock \emph{Advances in neural information processing systems}, pages
  115--132, 1999.

\bibitem[Huang et~al.(2005)Huang, Li, and Chen]{huang2005link}
Z.~Huang, X.~Li, and H.~Chen.
\newblock Link prediction approach to collaborative filtering.
\newblock In \emph{Proceedings of the 5th ACM/IEEE-CS joint conference on
  Digital libraries}, pages 141--142. ACM, 2005.

\bibitem[Kashima et~al.(2009)Kashima, Kato, Yamanishi, Sugiyama, and
  Tsuda]{kashima2009link}
H.~Kashima, T.~Kato, Y.~Yamanishi, M.~Sugiyama, and K.~Tsuda.
\newblock Link propagation: A fast semi-supervised learning algorithm for link
  prediction.
\newblock In \emph{SDM}, volume~9, pages 1099--1110. SIAM, 2009.

\bibitem[Kossinets and Watts(2006)]{kossinets2006empirical}
G.~Kossinets and D.J. Watts.
\newblock Empirical analysis of an evolving social network.
\newblock \emph{Science}, 311\penalty0 (5757):\penalty0 88--90, 2006.

\bibitem[Leskovec et~al.(2008)Leskovec, Backstrom, Kumar, and
  Tomkins]{leskovec2008microscopic}
J.~Leskovec, L.~Backstrom, R.~Kumar, and A.~Tomkins.
\newblock Microscopic evolution of social networks.
\newblock In \emph{Proceedings of the 14th ACM SIGKDD international conference
  on Knowledge discovery and data mining}, pages 462--470. ACM, 2008.

\bibitem[Liben-Nowell and Kleinberg(2007)]{liben2007link}
D.~Liben-Nowell and J.~Kleinberg.
\newblock The link-prediction problem for social networks.
\newblock \emph{Journal of the American society for information science and
  technology}, 58\penalty0 (7):\penalty0 1019--1031, 2007.

\bibitem[Lichtenwalter et~al.(2010)Lichtenwalter, Lussier, and
  Chawla]{lichtenwalter2010new}
R.N. Lichtenwalter, J.T. Lussier, and N.V. Chawla.
\newblock New perspectives and methods in link prediction.
\newblock In \emph{Proceedings of the 16th ACM SIGKDD international conference
  on Knowledge discovery and data mining}, pages 243--252. ACM, 2010.

\bibitem[Liu(2009)]{liu2009learning}
T.Y. Liu.
\newblock Learning to rank for information retrieval.
\newblock \emph{Foundations and Trends in Information Retrieval}, 3\penalty0
  (3):\penalty0 225--331, 2009.

\bibitem[Liu et~al.(2007)Liu, Liu, Qin, Ma, and Li]{liu2007supervised}
Y.T. Liu, T.Y. Liu, T.~Qin, Z.M. Ma, and H.~Li.
\newblock Supervised rank aggregation.
\newblock In \emph{Proceedings of the 16th international conference on World
  Wide Web}, pages 481--490. ACM, 2007.

\bibitem[L{\"u} and Zhou(2011)]{lu2011link}
L.~L{\"u} and T.~Zhou.
\newblock Link prediction in complex networks: A survey.
\newblock \emph{Physica A: Statistical Mechanics and its Applications},
  390\penalty0 (6):\penalty0 1150--1170, 2011.

\bibitem[Menon and Elkan(2011)]{menon2011link}
A.K. Menon and C.~Elkan.
\newblock Link prediction via matrix factorization.
\newblock In \emph{Machine Learning and Knowledge Discovery in Databases},
  pages 437--452. Springer, 2011.

\bibitem[Murata and Moriyasu(2007)]{murata2007link}
T.~Murata and S.~Moriyasu.
\newblock Link prediction of social networks based on weighted proximity
  measures.
\newblock In \emph{International Conference on Web Intelligence}, pages 85--88.
  IEEE, 2007.

\bibitem[Ngonmang et~al.(2012)Ngonmang, Viennet, and
  Tchuente]{ngonmang2012churn}
B.~Ngonmang, E.~Viennet, and M.~Tchuente.
\newblock Churn prediction in a real online social network using local
  community analysis.
\newblock In \emph{Proceedings of the 2012 International Conference on Advances
  in Social Networks Analysis and Mining (ASONAM 2012)}, pages 282--288. IEEE
  Computer Society, 2012.

\bibitem[Pavlov and Ichise(2007)]{pavlov2007finding}
M.~Pavlov and R.~Ichise.
\newblock Finding experts by link prediction in co-authorship networks.
\newblock \emph{FEWS}, 290:\penalty0 42--55, 2007.

\bibitem[Pedregosa et~al.(2011)Pedregosa, Varoquaux, Gramfort, Michel, Thirion,
  Grisel, Blondel, Prettenhofer, Weiss, Dubourg, Vanderplas, Passos,
  Cournapeau, Brucher, Perrot, and Duchesnay]{scikit-learn}
F.~Pedregosa, G.~Varoquaux, A.~Gramfort, V.~Michel, B.~Thirion, O.~Grisel,
  M.~Blondel, P.~Prettenhofer, R.~Weiss, V.~Dubourg, J.~Vanderplas, A.~Passos,
  D.~Cournapeau, M.~Brucher, M.~Perrot, and E.~Duchesnay.
\newblock Scikit-learn: Machine learning in {P}ython.
\newblock \emph{Journal of Machine Learning Research}, 12:\penalty0 2825--2830,
  2011.

\bibitem[Pujari and Kanawati(2012)]{pujari2012supervised}
M.~Pujari and R.~Kanawati.
\newblock Supervised rank aggregation approach for link prediction in complex
  networks.
\newblock In \emph{Proceedings of the 21st international conference companion
  on World Wide Web}, pages 1189--1196. ACM, 2012.

\bibitem[Raeder et~al.(2011)Raeder, Lizardo, Hachen, and
  Chawla]{raeder2011predictors}
T.~Raeder, O.~Lizardo, D.~Hachen, and N.V. Chawla.
\newblock Predictors of short-term decay of cell phone contacts in a large
  scale communication network.
\newblock \emph{Social Networks}, 33\penalty0 (4):\penalty0 245--257, 2011.

\bibitem[Scellato et~al.(2011)Scellato, Noulas, and
  Mascolo]{scellato2011exploiting}
S.~Scellato, A.~Noulas, and C.~Mascolo.
\newblock Exploiting place features in link prediction on location-based social
  networks.
\newblock In \emph{Proceedings of the 17th ACM SIGKDD international conference
  on Knowledge discovery and data mining}, pages 1046--1054. ACM, 2011.

\bibitem[Sculley(2007)]{sculley2007rank}
D.~Sculley.
\newblock Rank aggregation for similar items.
\newblock In \emph{SDM}, pages 587--592. SIAM, 2007.

\bibitem[Subbian and Melville(2011)]{subbian2011supervised}
K.~Subbian and P.~Melville.
\newblock Supervised rank aggregation for predicting influencers in twitter.
\newblock In \emph{Privacy, Security, Risk and Trust (PASSAT) and 2011 IEEE
  Third Inernational Conference on Social Computing (SocialCom), 2011 IEEE
  Third International Conference on}, pages 661--665. IEEE, 2011.

\bibitem[Tabourier et~al.(2019)Tabourier, Bernardes, Libert, and
  Lambiotte]{tabourier2019rankmerging}
Lionel Tabourier, Daniel~F Bernardes, Anne-Sophie Libert, and Renaud Lambiotte.
\newblock Rankmerging: a supervised learning-to-rank framework to predict links
  in large social networks.
\newblock \emph{Machine Learning}, 108:\penalty0 1729--1756, 2019.

\bibitem[Tylenda et~al.(2009)Tylenda, Angelova, and
  Bedathur]{tylenda2009towards}
T.~Tylenda, R.~Angelova, and S.~Bedathur.
\newblock Towards time-aware link prediction in evolving social networks.
\newblock In \emph{Proceedings of the 3rd Workshop on Social Network Mining and
  Analysis}, page~9. ACM, 2009.

\bibitem[Viswanath et~al.(2009)Viswanath, Mislove, Cha, and
  Gummadi]{viswanath-2009-activity}
Bimal Viswanath, Alan Mislove, Meeyoung Cha, and Krishna~P. Gummadi.
\newblock On the evolution of user interaction in facebook.
\newblock In \emph{Proceedings of the 2nd ACM SIGCOMM Workshop on Social
  Networks (WOSN'09)}, August 2009.

\bibitem[Yang et~al.(2015)Yang, Lichtenwalter, and Chawla]{yang2015evaluating}
Y.~Yang, R.N. Lichtenwalter, and N.V. Chawla.
\newblock Evaluating link prediction methods.
\newblock \emph{Knowledge and Information Systems}, 45\penalty0 (3):\penalty0
  751--782, 2015.

\bibitem[Zhou et~al.(2009)Zhou, L{\"u}, and Zhang]{zhou2009predicting}
T.~Zhou, L.~L{\"u}, and Y.C. Zhang.
\newblock Predicting missing links via local information.
\newblock \emph{The European Physical Journal B}, 71\penalty0 (4):\penalty0
  623--630, 2009.

\end{thebibliography}


\end{document}